\documentclass[a4paper,11pt]{article}
\pdfoutput=1
\usepackage{jheppub}

\usepackage{epsf}
\usepackage{epsfig}
\usepackage{subfigure}
\usepackage{mathtools}
\usepackage{hhline}
\usepackage{float}
\usepackage{multirow}
\usepackage{nicefrac}
\usepackage{epstopdf}
\usepackage{slashed}
\usepackage{xcolor}
\usepackage{url}
\usepackage{umoline} 
\allowdisplaybreaks

\newcommand{\dis}[1]{\begin{equation}\begin{split}#1\end{split}\end{equation}}

\newcommand{\be}{\begin{eqnarray}}
\newcommand{\ee}{\end{eqnarray}}

\newcommand{\ba}{\begin{array}}
\newcommand{\ea}{\end{array}}
\newcommand{\bee}{\begin{equation}\ba{c}}
\newcommand{\eee}{\ea\end{equation}}

\newcommand{\bi}{\begin{itemize}}
\newcommand{\ei}{\end{itemize}}

\toccontinuoustrue

\title{Hunting for Vectorlike Quarks}
\author{Radovan Derm\'i\v{s}ek$^1$,}
\author{Enrico Lunghi$^1$}
\author{and Seodong Shin$^{2,3}$}
\affiliation{
$^1$Physics Department, Indiana University, Bloomington, IN 47405, USA \\
$^2$Enrico Fermi Institute, University of Chicago, Chicago, IL 60637, USA \\
$^3$Department of Physics and IPAP, Yonsei University, Seoul 03722, Korea \\
}
\emailAdd{dermisek@indiana.edu} 
\emailAdd{elunghi@indiana.edu} 
\emailAdd{shinseod@indiana.edu}

\abstract{
We analyze decays of vectorlike quarks  in extensions of the standard model and a two Higgs doublet model. We identify several typical  patterns of branching ratios of the lightest new up-type quark, $t_4$, and down-type quark, $b_4$, depending on the structure of Yukawa couplings that mix the vectorlike and standard model quarks (we assume only mixing with the third generation) and also on their doublet or singlet nature. We find that decays  into heavy neutral or charged Higgs bosons, when kinematically open,  can easily dominate and even be close to 100\%:  $b_4\to H b$ at medium to large $\tan \beta$, $t_4\to H t$ at small  $\tan \beta$ and $b_4\to H^\pm t$, $t_4\to H^\pm b$ at both large and small $\tan \beta$. The pair production of vectorlike quarks  leads to $6 t$, $4t 2b$, $2t4b$ and $6b$ final states. The decay modes into $W$, $Z$ and $h$ follow the pattern expected from the Goldstone boson equivalence limit that we generalize to scenarios with all possible couplings. We also discuss in detail the structure of Yukawa couplings required to significantly deviate from the pattern characteristic of the Goldstone boson equivalence limit that can result in essentially arbitrary branching ratios.
\newpage 
}
\preprint{
\begin{minipage}{3cm}
\small
\flushright
EFI-19-1
\end{minipage}} 

\begin{document}

\maketitle 

\section{Introduction}
\label{sec:introduction}
Models with extra Higgs bosons or vectorlike quarks and leptons are among the simplest extensions of the standard model (SM). Consequently many strategies have been designed to search for them at collider experiments~\cite{Aaboud:2018xuw,Aaboud:2017zfn,Aaboud:2018pii,Sirunyan:2018fjh,Sirunyan:2017pks,Sirunyan:2018omb,Aaboud:2018saj,ATLAS:2018qxs,Sirunyan:2017ynj,Sirunyan:2018ncp,Sirunyan:2018rfo,Aaboud:2017sjh,Aaboud:2018gjj,Sirunyan:2018zut}. 
The success of this effort depends on understanding the decay patterns of new Higgs bosons or vectorlike matter. It is often assumed that only one particle or a specific coupling of a new particle is present. However, if two new particles are present, or more than one coupling of a new particle is sizable, the decay patterns can be dramatically altered. 

We analyze decays of new quarks  in extensions of the standard model and a two Higgs doublet model (type-II) by vectorlike pairs of new quarks (VLQ), corresponding to a copy of  SM quark SU(2) doublets and singlets and their vectorlike partners. We identify several typical  patterns of branching ratios of the lightest new up-type quark, $t_4$, and down-type quark, $b_4$, depending on the structure of Yukawa couplings that mix the vectorlike and standard model quarks  and also on their doublet or singlet nature. We assume only mixing with the third generation of SM quarks, nevertheless the results can be straightforwardly generalized for cases of mixing with the first or second generation. 

We find that decays  into heavy neutral ($H$) or charged ($H^\pm$) Higgs bosons, when kinematically open, can easily dominate and even be close to 100\%:  $b_4\to H b$ at medium to large $\tan \beta$, $t_4\to H t$ at small  $\tan \beta$ and $b_4\to H^\pm t$, $t_4\to H^\pm b$ at both large and small $\tan \beta$.
Thus, the pair production of vectorlike quarks  leads to $6 t$, $4t 2b$, $2t4b$ and $6b$ final states (and  similar final states for single production). The SM backgrounds for these final states (at large invariant mass)  are very small  and  
thus searching for these processes could lead to the simultaneous discovery of a new Higgs boson and a new quark. 

The usual decay modes into $W$, $Z$ and the SM Higgs boson, $h$, cluster around the pattern expected from the Goldstone boson equivalence limit (GBEL), corresponding to sending all vectorlike quark masses to infinity, that we generalize to scenarios with all possible couplings. For singlet-like new quarks this leads to 2:1:1 branching ratios into $W$, $Z$ and $h$. For doublet-like new quarks this leads to a one parameter family of branching ratios characterized by an arbitrary branching ratio to $W$ and equal branching ratios to $Z$ and $h$. We also discuss in detail the structure of Yukawa couplings required to significantly deviate from the pattern characteristic of  Goldstone boson equivalence limit  that can result in essentially arbitrary branching ratios. 

Extensions of the SM, two Higgs doublet models or the minimal supersymmetric model (MSSM) with vectorlike  matter were previously explored in a variety of contexts. Examples  include studies of their effects on gauge and Yukawa couplings in the framework of grand unification~\cite{Babu:1996zv, Kolda:1996ea, Ghilencea:1997yr, AmelinoCamelia:1998tm, BasteroGil:1999dx, Dermisek:2012as, Dermisek:2012ke, Dermisek:2017ihj, Dermisek:2018hxq} and on electroweak symmetry breaking and the Higgs boson mass~\cite{Babu:2008ge, Martin:2009bg, Dermisek:2016tzw}. The supersymmetric extension with a complete vectorlike family provides a very sharp prediction for the weak mixing angle~\cite{Moroi:1993, Dermisek:2017ihj} and also a possibility to understand the values of all large couplings in the SM  from the IR fixed point structure of the renormalization group equations~\cite{Dermisek:2018ujw}. In addition, vectorlike fermions are often introduced on purely phenomenological grounds to explain various anomalies. Examples include discrepancies in precision Z-pole observables~\cite{Choudhury:2001hs, Dermisek:2011xu, Dermisek:2012qx, Batell:2012ca} and the muon g-2 anomaly~\cite{Kannike:2011ng, Dermisek:2013gta, Dermisek:2014cia} among many others.

In this paper we focus on vectorlike quarks with the same quantum numbers as the quarks in the SM. Examples of signatures of related scenarios with vectorlike leptons can be found in Refs.~\cite{Dermisek:2015vra, Dermisek:2015oja, Dermisek:2015hue, Dermisek:2016via, Dermisek:2014qca, CidVidal:2018eel}.  For related studies and especially for studies of decay modes and signatures in scenarios with different quantum numbers of vectorlike matter see also Refs.~\cite{Cacciapaglia:2010vn, Okada:2012gy, Aguilar-Saavedra:2013qpa, Alok:2014yua, Banerjee:2016wls, Dobrescu:2016pda, Chala:2017xgc, Kim:2018mks, Das:2018gcr, Alhazmi:2018whk, Liu:2018hum}  and references therein.
 
This paper is organized as follows. In Sec.~\ref{sec:model},  we outline the model and assumptions. Details of the analysis and experimental constraints are discussed in Sec.~\ref{sec:scan}. The  main results and their discussion are contained in Sec.~\ref{sec:results} and we conclude in Sec.~\ref{sec:conclusions}. The appendix contains details of the model, formulas for couplings and relevant partial widths, and approximate formulas that are useful to understand the results.

\section{Model}
\label{sec:model}
We consider an extension of a two Higgs doublet model by vectorlike pairs of new quarks: SU(2) doublets $Q_{L,R}$ and SU(2) singlets $T_{L,R}$ and $B_{L,R}$. The quantum numbers of new particles are summarized in table~\ref{table:fieldcontents}. The $Q_L$, $T_R$ and $B_R$ have the same quantum numbers as the SM quark doublet $q_L$ and the right-handed quark singlets $u_R$ and $d_R$, respectively. We further assume that quarks couple to the two Higgs doublets as in the type-II model, namely the down sector couples to $H_d$ and the up sector couples to $H_u$. This can be achieved by the $Z_2$ symmetry specified in table \ref{table:fieldcontents}. The generalization to the whole vectorlike family of new fermions, including the lepton sector (which has been studied in ref.~\cite{Dermisek:2015oja}), is straightforward. 
\begin{table}[htp]
\begin{center}
\begin{tabular}{c c c c c c c c c}
\hline
\hline
 & ~~$q^i_L$ & ~~$u^i_R$ & ~~$d^i_R$ & ~~$Q_{L,R}$ & ~~$T_{L,R}$ & ~~$B_{L,R}$ & ~~$H_d$ & ~~ $H_u$\\
\hline
SU(2)$_{\rm L}$ & ~~\bf 2 & ~~\bf 1 & ~~\bf 1 & ~~\bf 2 & ~~\bf 1 & ~~\bf 1 & ~~\bf 2 & ~~\bf 2 \\
U(1)$_{\rm Y}$ & ~~$\frac16$ & ~~$\frac23$ & ~~-$\frac13$ & ~~$\frac16$ & ~~$\frac23$ & ~~-$\frac13$ & ~~$\frac12$ & ~~-$\frac12$ \\
Z$_2$ & ~~+ & ~~+ & ~~-- & ~~+ & ~~+ & ~~-- & ~~-- & ~~+ \\
\hline
\hline
\end{tabular}
\end{center}
\caption{Quantum numbers of standard model quarks ($q^i_L, u^i_R, d^i_R$ for $i=1,2,3$), extra vectorlike quarks and the two Higgs doublets. The electric charge is given by $Q = T_3 +Y$, where $T_3$ is the weak isospin, which is +1/2 for the first component of a doublet and -1/2 for the second component. }
\label{table:fieldcontents}
\end{table}
The most general renormalizable Lagrangian consistent with our assumptions contains the following Yukawa and mass terms for the SM and vectorlike quarks:
\dis{
{\cal L} \supset \;  & - y_d^{ij} \bar q^i_L  d^j_R H_d - \lambda_B^i \bar q^i_L  B_R H_d  -  \lambda_Q^j \bar Q_L d^j_R H_d -  \lambda \bar Q_L  B_R H_d - \bar \lambda H_d^\dagger \bar B_L  Q_R  \\
& - y_u^{ij} \bar q^i_L  u^j_R H_u - \kappa_T^i  \bar q^i_L  T_R H_u - \kappa_Q^j  \bar Q_L  u^j_R H_u - \kappa  \bar{Q}_L  T_R H_u - \bar \kappa H_u^\dagger \bar{T}_L Q_R  \\
&  - M_Q \bar Q_L Q_R - M_T \bar T_L T_R - M_B \bar B_L B_R + {\rm h.c.}~,
\label{eq:lagrangian}
}
where the first term represents the Yukawa couplings of the SM down-type quarks, followed by Yukawa couplings of vectorlike quarks to $H_d$ (denoted by various $\lambda$s), Yukawa couplings of the SM up-type quarks, Yukawa couplings of vectorlike quarks to $H_u$ (denoted by various $\kappa$s), and finally by mass terms for vectorlike quarks. 
Note that the explicit mass terms mixing SM and vectorlike quarks, $M^i_{Q} \bar q_L^i Q_R$, $M^i_{T} \bar T_L u_R^i$ and $M^i_{B} \bar B_L d_R^i$, can be removed by redefinitions of $Q_L$, $T_R$, $B_R$ and the Yukawa couplings. 
The components of doublets are labeled as follows:
\dis{
q^i_L  = \left( 
\begin{array}{c}
u^i_L \\
d^i_L
\end{array}
\right),~
Q_{L,R}  = \left( 
\begin{array}{c}
T_{L,R}^Q \\
B_{L,R}^Q
\end{array}
\right),~
H_d  = \left( 
\begin{array}{c}
H_d^+ \\
H_d^0 
\end{array}
\right),~
H_u  = \left( 
\begin{array}{c}
H_u^0\\
H_u^- 
\end{array}
\right).
}
We assume that the neutral Higgs components develop real and positive vacuum expectation values, $\left< H_u^0 \right> = v_u$ and $\left< H_d^0 \right> = v_d$, as in the $CP$ conserving two Higgs doublet model with $\sqrt{v_u^2 + v_d^2} = v = 174$ GeV and we define $\tan \beta \equiv v_u / v_d$. 

For simplicity, we further assume that the new quarks mix only with one family of SM quarks and we consider the mixing with the third family as an example; the  mixing of new quarks with more than one SM family simultaneously is strongly constrained by various flavor changing processes and we will not pursue this direction here. In the basis in which the SM quark Yukawas are diagonal, the mass matrices describing the mixing between the third generation and the vectorlike quarks are (see the appendix):
\begin{align}
\left( \begin{array}{ccc} \bar t_L & \bar T_L^Q & \bar T_L \end{array}\right)
M_t 
\left( \begin{array}{c}t_R  \\T_R^Q \\T_R\end{array} \right) 
&= 
\left( \begin{array}{ccc}\bar t_L & \bar T_L^Q& \bar T_L\end{array}\right)
\left( \begin{array}{ccc}
y_t v_u & 0 & \kappa_T v_u \\
\kappa_Q  v_u & M_Q & \kappa v_u \\
0 & \bar \kappa v_u & M_T \\
\end{array}\right)
\left(\begin{array}{c}t_R  \\T_R^Q \\T_R\end{array}\right)~,
\label{eq:mmu} \\
\left( \begin{array}{ccc}\bar b_L & \bar B_L^Q & \bar B_L\end{array}\right)
M_b
\left( \begin{array}{c}b_R  \\B_R^Q \\B_R\end{array}\right) 
&= 
\left( \begin{array}{ccc}\bar b_L & \bar B_L^Q & \bar B_L\end{array}\right)
\left( \begin{array}{ccc}
y_b v_d & 0 & \lambda_B v_d \\
\lambda_Q v_d & M_Q & \lambda v_d \\
0 & \bar \lambda v_d & M_B \\
\end{array}\right)
\left(\begin{array}{c}b_R  \\B_R^Q \\B_R\end{array}\right)~.
\label{eq:mmd}
\end{align}
Note that the corresponding mass matrices in the case of a single Higgs doublet can be obtained by setting $v_u=v_d=174\; {\rm GeV}$. A complete discussion of the mass eigenstates and of their couplings to the $W$, $Z$, and Higgs bosons can be found in the appendix.

\section{Parameter space scan and experimental constraints}
\label{sec:scan}
We study the branching ratio patterns that can be obtained in the model by varying the relevant parameters as follows:
\begin{align}
M_{Q,T,B} &\in [900, 4000] {\rm GeV} \; , \\
\kappa_T, \kappa_Q, \kappa, \bar \kappa &\in [-1.0, 1.0]~({\rm if\, mixing\, exists}) \; , \\
\lambda_B, \lambda_Q, \lambda, \bar \lambda &\in [-1.0, 1.0]~({\rm if\, mixing\, exists}) \; , \\
\tan\beta & \in [0.3, 50] \; . 
\end{align}
Note that the upper range of the couplings has no impact on presented results as long as it is common for all couplings.

We impose the experimental constraints from precision electroweak measurements~\cite{Tanabashi:2018oca}\footnote{See also Ref.~\cite{Chen:2017hak} for a detailed discussion about imposing constraints from precision  electroweak  measurements in models with vectorlike quarks.}, $h \to (\gamma \gamma,4\ell)$~\cite{ATLAS:2017myr, ATLAS:2018doi}\footnote{Constraints  from related experimental results~\cite{Khachatryan:2014ira,Aaboud:2018xdt,Sirunyan:2018aui} are not very different from constraints resulting from Ref.~\cite{ATLAS:2017myr}.} and direct searches for vectorlike quarks pair produced at the LHC~\cite{Aaboud:2017zfn,Aaboud:2018pii,Aaboud:2018xuw,ATLAS:2018iwl}.
For the latter constraints, we directly use the data points from hepdata.net~\cite{hepdata:vlqpair} obtained from Ref.~\cite{Aaboud:2017zfn} where the limits in terms of mass of vectorlike quarks and the branching ratios into $W,Z$ and the SM Higgs boson are similar to the other search results.

Note that searches for the single production of VLQ via $t$-channel quark gluon interactions also exist but the experimental constraints are not stronger than those for the pair production of VLQ~\cite{Sirunyan:2017tfc,ATLAS:2018qxs} although the production cross section can be larger than that of the pair production for VLQ heavier than $\sim 800$ GeV~\cite{Aguilar-Saavedra:2013qpa}.
Hence, we do not consider the corresponding experimental bounds here but leave it to a future work~\cite{future1}.
In scenarios where the lightest VLQ can decay into heavy neutral or charged Higgs boson, the currently existing searches for VLQ, focussed on decay modes into $W$, $Z$, and $h$, do not directly apply.

In the model we are considering we expect potentially large contributions to $B\to X_s \gamma$. While the charged Higgs loop is suppressed for Higgs masses above 1 TeV, the $b_R W t_R$ vertex given in eq.~(\ref{gWtRbR}) yields a diagram which is chirally enhanced. Note that this vertex requires couplings to both $H_u$ and $H_d$. This diagram yields a contribution to the coefficient of the magnetic moment operator $C_7$ (see, for instance, ref.~\cite{Buras:1998raa}) which is enhanced by a factor $\frac{m_t}{m_b} \frac{v_u v_d}{M_Q^2} \kappa_Q \lambda_Q \sim \frac{1}{2} \sin2 \beta \; \kappa_Q \lambda_Q \frac{1 \text{TeV}^2}{M_Q^2}$. This contribution is unacceptably large for moderate $\tan\beta$ and $\kappa_Q \lambda_Q \gtrsim 10^{-2}$. 

Contributions to other flavor transitions are much smaller because, in absence of chiral enhancements, experiments constrain various combinations of the entries of the effective CKM matrix given in eq.~(\ref{eq:CKM}). Any deviation from the SM expectation is related to the non-unitarity of this matrix, which is of order $v_u^2 \kappa_T^2/2M_T^2$ or $v_d^2 \lambda_B^2/2M_B^2$, yielding contributions which can be at most 1.5\% for vectorlike quarks heavier than 1 TeV. 

Finally we note that in our framework there are no flavor changing neutral interactions in the Higgs nor $Z$ sector. Therefore,  $\tan\beta$ enhanced tree-level Higgs contributions to $B_{d,s} \to \ell^+\ell^-$, which are present in supersymmetric models with loop-induced non-holomorphic Higgs couplings, are absent.

\begin{table}[t]
\begin{center}
\begin{tabular}{|c|c|c|} \hline
\small $ \frac{64\pi}{m_{t_4}}\, \Gamma$  & \small doublet & \small singlet \\\hline\hline
\small  $t_4 \to W^+ b$ & $2 \lambda_Q^2 c^2_\beta$ & $2 \kappa_T^2 s^2_\beta$ \\ \hline
\small  $t_4 \to Z t$ & $\kappa_Q^2 s^2_\beta$ & $\kappa_T^2 s_\beta^2$   \\ \hline
\small  $t_4 \to h t$ & $\kappa_Q^2 s^2_\beta$ & $\kappa_T^2 s_\beta^2$   \\ \hline
\small  $t_4 \to H t$ & $\kappa_Q^2 c^2_\beta$ & $\kappa_T^2 c_\beta^2$   \\ \hline
\small  $t_4 \to H^+ b$ & $  2\lambda_Q^2 s^2_\beta$&  $  2\kappa_T^2 c^2_\beta$   \\ \hline
\end{tabular}
\;\;
\begin{tabular}{|c|c|c|} \hline
\small $ \frac{64\pi}{m_{b_4}}\, \Gamma$ & \small doublet & \small singlet \\\hline\hline
\small  $b_4 \to W^+ t$ & $2 \kappa_Q^2 s^2_\beta$ & $2 \lambda_B^2 c^2_\beta$   \\ \hline
\small  $b_4 \to Z b$ & $\lambda_Q^2 c^2_\beta$  &  $\lambda_B^2 c^2_\beta$  \\ \hline
\small  $b_4 \to h b$ & $\lambda_Q^2 c^2_\beta$& $\lambda_B^2 c^2_\beta$   \\ \hline
\small  $b_4 \to H b$ & $\lambda_Q^2 s^2_\beta$ &  $\lambda_B^2 s^2_\beta$  \\ \hline
\small  $b_4 \to H^+ t$ & $  2\kappa_Q^2 c^2_\beta$ & $  2\lambda_B^2 s^2_\beta$   \\ \hline
\end{tabular}
\end{center}
\caption{Leading dependence of the $t_4$ and $b_4$ decay widths on the Lagrangian parameters in the large mass limit. We use the shorthand notation $s_\beta = \sin\beta$ and $c_\beta = \cos \beta$.  \label{tab:couplings}}
\end{table}
%


\section{Results}
\label{sec:results}
The main features of the decays of vectorlike quarks can be understood from the dominant couplings that appear in table~\ref{tab:couplings}, which can be easily read from the approximate expressions presented in the appendix. An important quantity that controls the decay of the lightest vectorlike quark is its doublet or singlet fraction, which add up to unity and, for up-type vectorlike quarks $t_a$ ($a=4,5$), are defined respectively as 
\begin{align}
d_{t_a} &= \frac12 \left\{ [(V_L^u)^\dagger_{a3}]^2 + [(V_L^u)^\dagger_{a4}]^2 + [(V_R^u)^\dagger_{a4}]^2  \right\}\, , 
\label{eq:doubletfr} \\
s_{t_a} &= \frac12 \left\{ [(V_R^u)^\dagger_{a3}]^2 + [(V_L^u)^\dagger_{a5}]^2 + [(V_R^u)^\dagger_{a5}]^2   \right\}\, .
\label{eq:singletfr} 
\end{align}
The doublet and singlet fractions for down-type vectorlike quarks $b_a$ ($a=4,5$) are obtained by replacement $V_{L,R}^u \to V_{L,R}^d$. Note that in the absence of couplings to $H_u$ ($H_d$), i.e., in the case when all $\lambda$'s ($\kappa$'s) in \eqref{eq:lagrangian} vanish, the doublet fraction of $t_4$ ($b_4$) defined in eq.~(\ref{eq:doubletfr}) is exactly 1 or 0. The latter case corresponds to a vectorlike singlet quark which is completely uncoupled (and thus of no interest for our study). 

\begin{table}
\begin{center}
\begin{tabular}{|c|c|c|c|c|c|c|} \hline
\multirow{2}{*}{} & \multicolumn{2}{c|}{\small Couplings to $H_u$ only} & \multicolumn{2}{c|}{\small Couplings to $H_d$ only} & \multicolumn{2}{c|}{\small Couplings to $H_u$ and $H_d$}\\ \cline{2-7}
 & \small doublet & \small singlet & \small doublet & \small singlet & \small doublet & \small singlet\\\hline\hline
 \small ${\rm BR}(t_4 \to W^+ b)$ & 0 & \small $1/2$ & 1 & 0 &\small  ${\scriptstyle (1-x)} \frac{1}{1+x(t_\beta^2-1)}$ & \small $1/2$\\ \hline
 \small  ${\rm BR}(t_4 \to Z t)$ & \small $1/2$ & \small $1/4$ & 0 & 0& \small  $\frac{x}{2}\frac{t_\beta^2}{1+x(t_\beta^2-1)}$ & \small $1/4$\\\hline
 \small  ${\rm BR}(t_4 \to h t)$ & \small $1/2$ &\small  $1/4$ & 0 & 0& \small  $\frac{x}{2}\frac{t_\beta^2}{1+x(t_\beta^2-1)}$ &\small $1/4$ \\ \hline\hline
 \small  ${\rm BR}(b_4 \to W^+ t)$ & 1 & 0 & 0 &\small  $1/2$ &\small   ${\scriptstyle x}\frac{t_\beta^2}{1+x(t_\beta^2-1)}$ & $1/2$ \\ \hline
 \small  ${\rm BR}(b_4 \to Z b)$ & 0 & 0 & \small $1/2$ &\small  $1/4$ &\small   $\frac{1-x}{2}\frac{1}{1+x(t_\beta^2-1)}$ &\small  $1/4$\\ \hline
 \small  ${\rm BR}(b_4 \to h b)$ & 0 & 0 & \small $1/2$ & \small $1/4$  &\small  $\frac{1-x}{2}\frac{1}{1+x(t_\beta^2-1)}$ & \small $1/4$\\ \hline
\end{tabular}
\end{center}
\caption{Vectorlike quark branching ratios in the Goldstone boson equivalence limit. When couplings to both $H_u$ and $H_d$ are present the branching ratios are functions of $t_\beta = \tan\beta$ and $x=\kappa_Q^2/(\kappa_Q^2+\lambda_Q^2)$.
\label{tab:goldstone}}
\end{table}

An important point is that decays into heavy Higgses can easily dominate when kinematically open. A second point is that, when these decay channels are not open, vectorlike quarks branching ratios into $W$, $Z$ and $h$ follow a very simple pattern which we summarize in table~\ref{tab:goldstone}. If only couplings to $H_u$ ($H_d$) are present, the $t_4$ ($b_4$) branching ratios into $W$, $Z$ and $h$ are $(0,1/2,1/2)$ and $(1/2,1/4,1/4)$ for the doublet and singlet case respectively. If couplings to $H_u$ and $H_d$ are present simultaneously, the doublet branching ratios into $Z$ and $h$ are still equal but the $W$ channel can be as large as 100\%. The actual values of the branching ratios depend on $\tan\beta$ and on the relative size of the $H_u$ and $H_d$ couplings which we parameterize as $x=\kappa_Q^2/(\kappa_Q^2+\lambda_Q^2)$, ($0\leq x\leq 1$).

In the next two subsections we discuss decays into heavy Higgses and then dwell into the intricacies of standard decays into $W$, $Z$ and $h$ in situations in which subdominant couplings become important.

\subsection{Decays into heavy Higgses}

\begin{figure}[t]
\begin{center}
\includegraphics[width=.49\linewidth]{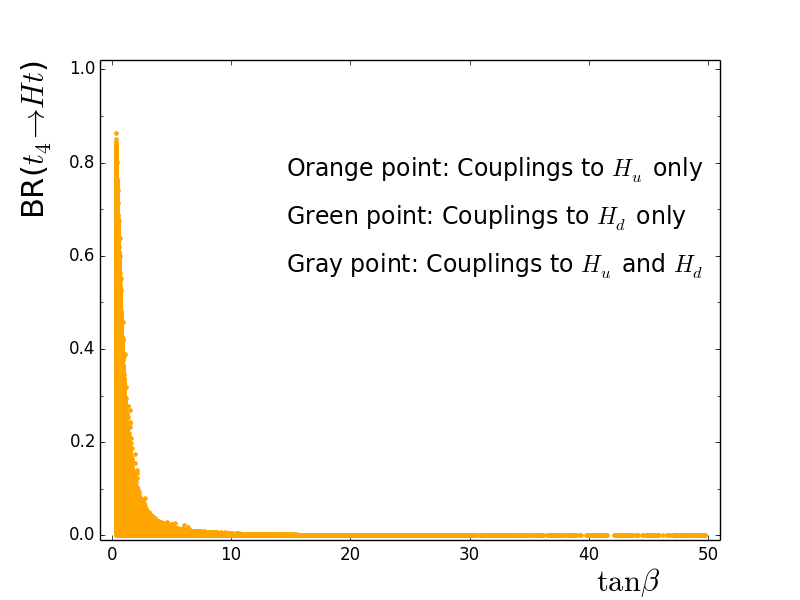} 
\includegraphics[width=.49\linewidth]{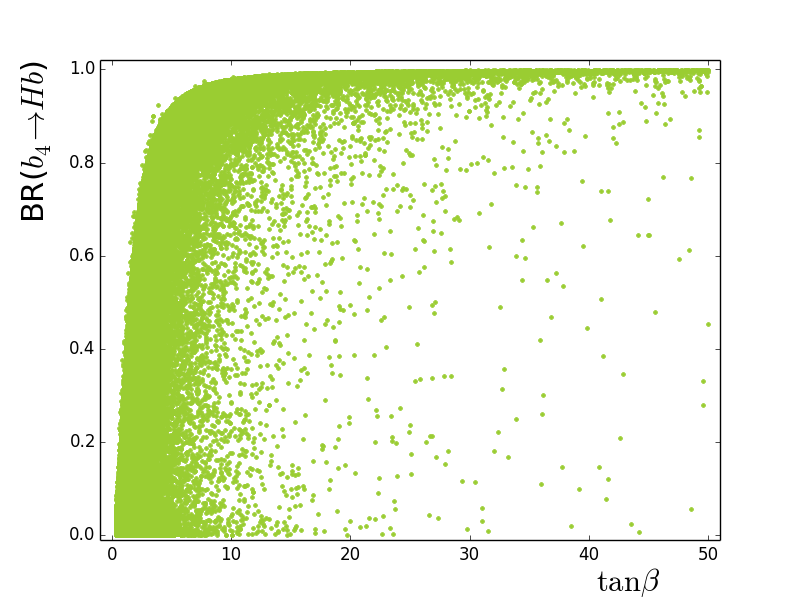} 
\includegraphics[width=.49\linewidth]{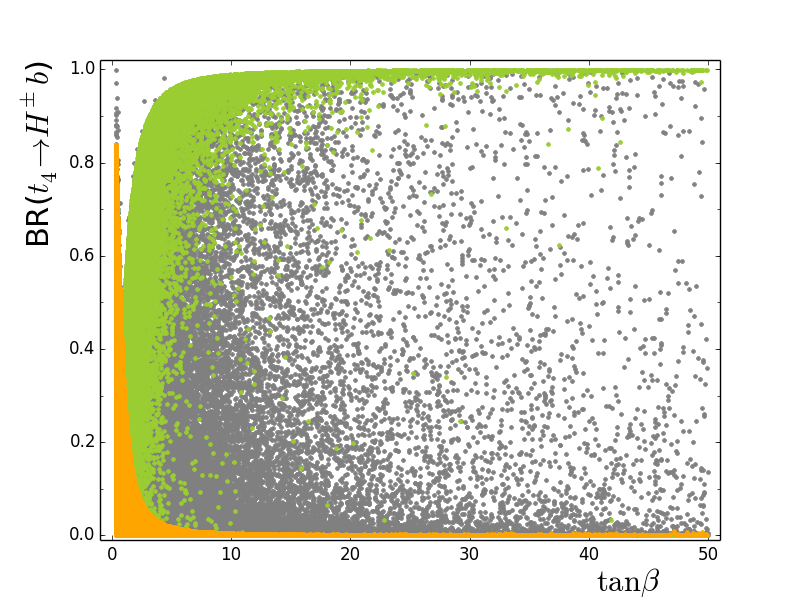}
\includegraphics[width=.49\linewidth]{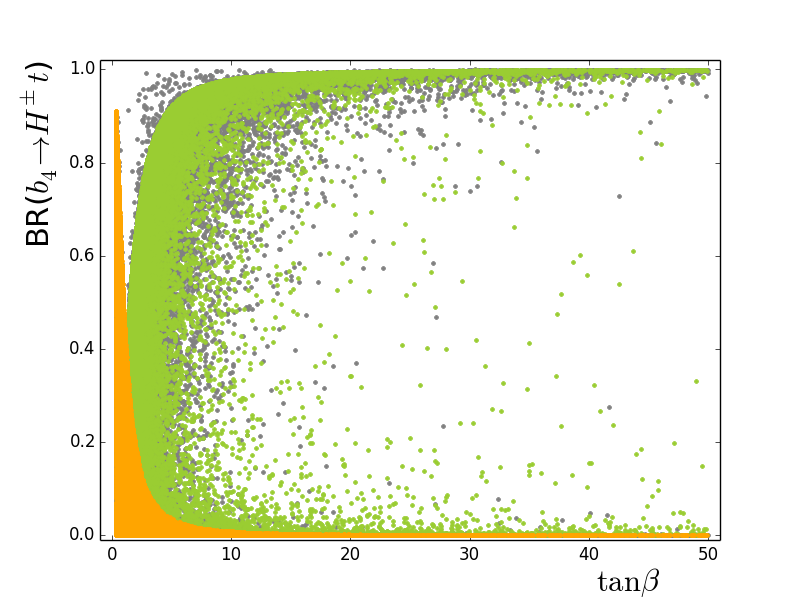} 
\caption{Allowed branching ratios of $t_4$ and $b_4$ into heavy charged and neutral Higgses. In all panels, orange (green) points correspond to couplings to $H_u$ ($H_d$) only; gray points are the additional possibilities allowed by a full parameter space scan.}
\label{fig:BRscenarios_intoH}
\end{center}
\end{figure}

In figure~\ref{fig:BRscenarios_intoH} we present the branching ratios of vectorlike quarks into heavy CP--even neutral and charged Higgses that we obtain from the parameter space scan described in section~\ref{sec:scan}. For simplicity, we assume that only one (either charged or neutral heavy Higgs) decay channel is kinematically open.\footnote{Note that the results for the decay mode into CP--odd neutral Higgs would be similar.} 
In all panels orange (green) points correspond to allowing only couplings to $H_u$ ($H_d$); gray points show the additional branching ratio values that we obtain by allowing simultaneous couplings to $H_u$ and $H_d$. 

The interpretation of these plots follows immediately from the couplings presented in table~\ref{tab:couplings} which yield the branching ratios collected in tables~\ref{tab:H0BR}-\ref{tab:BRratios}. 
The channel $t_4\to Ht$ is allowed only if couplings to $H_u$ are present and is sizable only at small $\tan\beta$ independently of the $t_4$ doublet fraction. Conversely, the channel $b_4\to H b$ requires couplings to $H_d$ and can easily dominate at medium-to-large $\tan\beta$ independently of the $b_4$ doublet fraction. 

For the decay $t_4\to H^\pm b$, couplings to $H_u$ ($H_d$) only, result in a mostly singlet (doublet) $t_4$ with large branching ratio at small (medium-to-large) $\tan\beta$. Similarly, for the decay $b_4\to H^\pm t$, couplings to $H_u$ ($H_d$) only, result in a mostly doublet (singlet) $b_4$ with large branching ratio at small (medium-to-large) $\tan\beta$. These features can be inferred directly from tables \ref{tab:couplings} and \ref{tab:BRratios}. Inspection of the bottom panels of figure~\ref{fig:BRscenarios_intoH} reveals also that, in presence of coupling to $H_d$ only and at large $\tan\beta$, the $t_4\to H^+ b$ branching ratio is typically close to unity while the $b_4\to H^- t$ one can be small. This happens because in presence of couplings to $H_d$ only, $t_4$ is either 100\% doublet or 100\% singlet (the doublet fraction in eq.~(\ref{eq:doubletfr}) is unity but it is possible to have a singlet $t_4$ by lowering $M_T$) while the $b_4$ doublet fraction varies continuously between 0 and 100\%. This implies that the $t_4\to H^+ b$ branching ratio is either close to 100\% (doublet $t_4$) or close to 0 (singlet $t_4$), while the $b_4\to H^- t$ one covers smoothly the entire range, with large (small) branching ratios corresponding to a mostly singlet (doublet) $b_4$. The presence of couplings to both $H_u$ and $H_d$ (gray points) allows the $t_4\to H^\pm b$ branching ratio to acquire any value by tuning the contributions of the two sets of couplings.

\begin{table}
\begin{center}
\begin{tabular}{|c|c|c|c|c|c|c|} \hline
\multirow{2}{*}{} & \multicolumn{2}{c|}{Couplings to $H_u$ only} & \multicolumn{2}{c|}{Couplings to $H_d$ only} & \multicolumn{2}{c|}{Couplings to $H_u$ and $H_d$}\\ \cline{2-7}
 & \small doublet & \small singlet & \small doublet & \small singlet & \small doublet & \small singlet\\\hline\hline
\small  ${\rm BR}(t_4 \to W^+ b)$ & 
0 & 
$\frac{2 t^2_\beta}{4 t^2_\beta+1}$ & 
1 & 
0 & 
\small  $\frac{2(1-x)}{2+x(2t_\beta^2-1)}$ & 
$\frac{2 t^2_\beta}{4 t^2_\beta+1}$
\\ \hline
\small  ${\rm BR}(t_4 \to Z t)$ & 
$\frac{t^2_\beta}{2 t^2_\beta+1}$ & 
$\frac{t^2_\beta}{4 t^2_\beta+1}$ & 
0 & 
0 &
\small  $\frac{x t_\beta^2}{2+x(2t_\beta^2-1)}$ & 
$\frac{t^2_\beta}{4 t^2_\beta+1}$
\\ \hline
\small  ${\rm BR}(t_4 \to h t)$ & 
$\frac{t^2_\beta}{2 t^2_\beta+1}$ & 
$\frac{t^2_\beta}{4 t^2_\beta+1}$ & 
0 & 
0 &
\small  $\frac{x t_\beta^2}{2+x(2t_\beta^2-1)}$ & 
$\frac{t^2_\beta}{4 t^2_\beta+1}$
\\ \hline
\small  ${\rm BR}(t_4 \to H t)$ & 
$\frac{1}{2 t^2_\beta+1}$ & 
$\frac{1}{4 t^2_\beta+1}$ & 
0 & 
0 &
\small  $\frac{x }{2+x(2t_\beta^2-1)}$& 
$\frac{1}{4 t^2_\beta+1}$
\\ \hline\hline
\small  ${\rm BR}(b_4 \to W^+ t)$ & 
1 & 
0 & 
0 & 
$\frac{2}{t^2_\beta+4}$ & 
\small  $\frac{2 x t_\beta^2}{t_\beta^2+2 + x (t_\beta^2-2)}$  & 
$\frac{2}{t^2_\beta+4}$ 
\\ \hline
\small  ${\rm BR}(b_4 \to Z b)$ & 
0 & 
0 & 
$\frac{1}{t^2_\beta+2}$ & 
$\frac{1}{t^2_\beta+4}$ &
\small  $\frac{1- x }{t_\beta^2+2 + x (t_\beta^2-2)}$  & 
$\frac{1}{t^2_\beta+4}$
\\ \hline
\small  ${\rm BR}(b_4 \to h b)$ & 
0 & 
0 & 
$\frac{1}{t^2_\beta+2}$ & 
$\frac{1}{t^2_\beta+4}$ & 
\small  $\frac{1- x }{t_\beta^2+2 + x (t_\beta^2-2)}$ & 
$\frac{1}{t^2_\beta+4}$
\\ \hline
\small  ${\rm BR}(b_4 \to H b)$ & 
0 & 
0 & 
$\frac{t^2_\beta}{t^2_\beta+2}$ & 
$\frac{t^2_\beta}{t^2_\beta+4}$ &
\small  $\frac{(1-x) t_\beta^2}{t_\beta^2+2 + x (t_\beta^2-2)}$ & 
$\frac{t^2_\beta}{t^2_\beta+4}$ 
\\ \hline
\end{tabular}
\end{center}
\caption{Branching ratios of the vectorlike quarks $t_4$ and $b_4$ into $W$, $Z$, $h$ and $H$ as a function of $t_\beta  = \tan\beta$  and $x=\kappa_Q^2/(\kappa_Q^2+\lambda_Q^2)$. These expressions are an excellent approximation for heavy vectorlike masses and in the limit in which the couplings in table~\ref{tab:couplings} dominate. \label{tab:H0BR}}
\end{table}
\begin{table}
\begin{center}
\begin{tabular}{|c|c|c|c|c|c|c|} \hline
\multirow{2}{*}{} & \multicolumn{2}{c|}{\small Couplings to $H_u$ only} & \multicolumn{2}{c|}{\small Couplings to $H_d$ only} & \multicolumn{2}{c|}{\small Couplings to $H_u$ and $H_d$}\\ \cline{2-7}
 & \small doublet & \small singlet & \small doublet & \small singlet & \small doublet & \small singlet\\\hline\hline
\small  ${\rm BR}(t_4 \to W^+ b)$ & 
0 & 
$\frac{2 t^2_\beta}{4 t^2_\beta+ 2}$ & 
$\frac{ 1}{t^2_\beta+ 1}$ & 
0 & 
 $\frac{1-x}{1-x+t_\beta^2}$ & 
$\frac{2 t^2_\beta}{4 t^2_\beta+ 2}$
\\ \hline
\small  ${\rm BR}(t_4 \to Z t)$  & 
\small 1/2 & $\frac{ t^2_\beta}{4 t^2_\beta+ 2}$ & 
0 & 
0 & 
 $\frac{x}{2}\frac{t_\beta^2}{1-x+t_\beta^2}$ & 
$\frac{ t^2_\beta}{4 t^2_\beta+ 2}$ 
\\ \hline
\small  ${\rm BR}(t_4 \to h t)$   & 
\small 1/2 & $\frac{ t^2_\beta}{4 t^2_\beta+ 2}$ & 
0 & 
0 & 
 $\frac{x}{2}\frac{t_\beta^2}{1-x+t_\beta^2}$ & 
$\frac{ t^2_\beta}{4 t^2_\beta+ 2}$  
\\ \hline
\small  ${\rm BR}(t_4 \to H^+ b)$ & 
0 & 
$\frac{ 2}{4 t^2_\beta+ 2}$ & 
$\frac{t^2_\beta}{t^2_\beta+ 1}$ & 
0  & 
 $\frac{(1-x)t_\beta^2}{1-x+t_\beta^2}$ & 
$\frac{  2}{4 t^2_\beta+ 2}$  
\\ \hline\hline
\small  ${\rm BR}(b_4 \to W^+ t)$ & 
$\frac{  t^2_\beta}{  t^2_\beta+1}$ & 
0 & 
0 & 
$\frac{2}{  2 t^2_\beta+4}$ & 
 $\frac{x t_\beta^2}{1+x t_\beta^2}$ & 
$\frac{ 2}{ 2 t^2_\beta+4}$ 
\\ \hline
\small  ${\rm BR}(b_4 \to Z b)$   & 
0 & 
0 & 
\small 1/2 & 
$\frac{ 1}{ 2 t^2_\beta+4}$ & 
 $\frac{1-x}{2}\frac{1}{1+x t_\beta^2}$ & 
$\frac{ 1}{  2t^2_\beta+4}$ 
\\ \hline
\small  ${\rm BR}(b_4 \to h b)$   & 
0 & 
0 & 
\small 1/2 & 
$\frac{ 1}{  2 t^2_\beta+4}$ & 
 $\frac{1-x}{2}\frac{1}{1+x t_\beta^2}$ & 
$\frac{ 1}{ 2 t^2_\beta+4}$ 
\\ \hline
\small  ${\rm BR}(b_4 \to H^+ t)$ & 
$\frac{1}{  t^2_\beta+1}$ & 
0 & 
0 & 
$\frac{  2 t^2_\beta}{  2 t^2_\beta+4}$ & 
 $\frac{x}{1+x t_\beta^2}$ & 
$\frac{  2 t^2_\beta}{  2 t^2_\beta+4}$ \\ \hline
\end{tabular}
\end{center}
\caption{Branching ratios of the vectorlike quarks $t_4$ and $b_4$ into $W$, $Z$, $h$ and $H^\pm$ as a function of $t_\beta  = \tan\beta$  and $x=\kappa_Q^2/(\kappa_Q^2+\lambda_Q^2)$. These expressions are an excellent approximation for heavy vectorlike masses and in the limit in which the couplings in table~\ref{tab:couplings} dominate. \label{tab:HPBR}}
\end{table}
\begin{table}
\begin{center}
\small
\begin{tabular}{|c|c|c|c|c|} \hline
\multirow{2}{*}{} & \multicolumn{2}{c|}{Couplings to $H_u$ only} & \multicolumn{2}{c|}{Couplings to $H_d$ only} \\ \cline{2-5}
 & doublet & singlet & doublet & singlet \\\hline
${\rm BR}(t_4 \to H t)/{\rm BR}(t_4 \to \text{SM})$ & $1/(2t^2_\beta)$ & $1/(4t^2_\beta)$ & 0 & 0 \\ \hline
${\rm BR}(b_4 \to H b)/{\rm BR}(b_4 \to \text{SM})$ & 0 & 0 & $t^2_\beta/2$ & $t^2_\beta/4$  \\ \hline
${\rm BR}(t_4 \to H^+ b)/{\rm BR}(t_4 \to \text{SM})$ &  0 & $1/(4t^2_\beta)$ & $t^2_\beta/2$ & 0 \\ \hline
${\rm BR}(b_4 \to H^+ t)/{\rm BR}(b_4 \to \text{SM})$ & $1/(2t^2_\beta)$ & 0 & 0 & $t^2_\beta/4$  \\ \hline
\end{tabular}
\end{center}
\caption{Branching ratios of the vectorlike quarks $(t_4,b_4)$ into $(H,H^\pm)$ normalized to the total branching ratio into SM particles ($W$, $Z$ and $h$) as a function of $t_\beta  = \tan\beta$. These expressions are an excellent approximation for heavy vectorlike masses and in the limit in which the couplings in table~\ref{tab:couplings} dominate. \label{tab:BRratios}}
\end{table}
\subsection{Decays into $W$, $Z$ and $h$}

\begin{figure}
\begin{center}
\includegraphics[width=.49\linewidth]{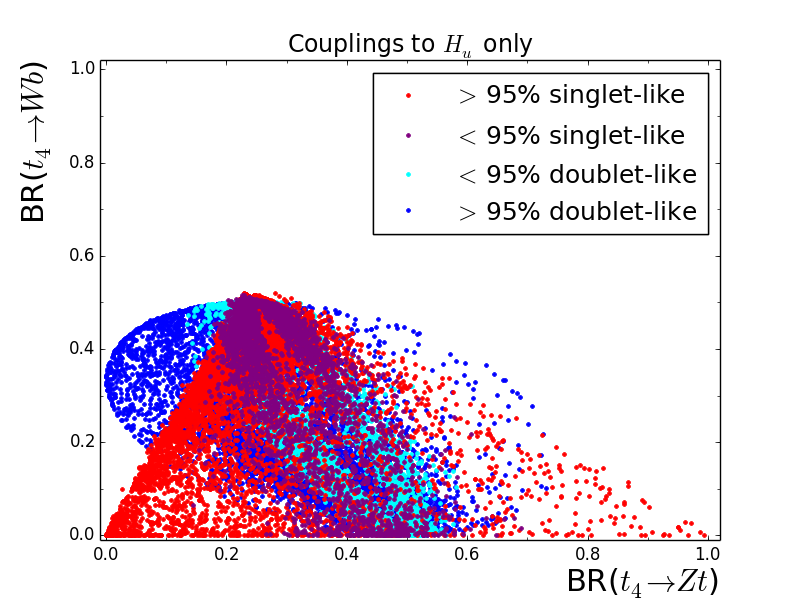} 
\includegraphics[width=.49\linewidth]{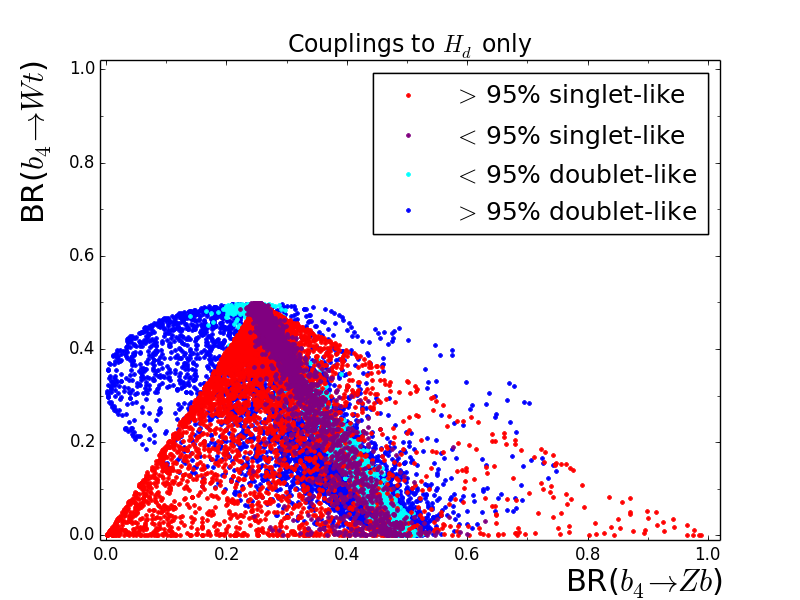} 
\includegraphics[width=.49\linewidth]{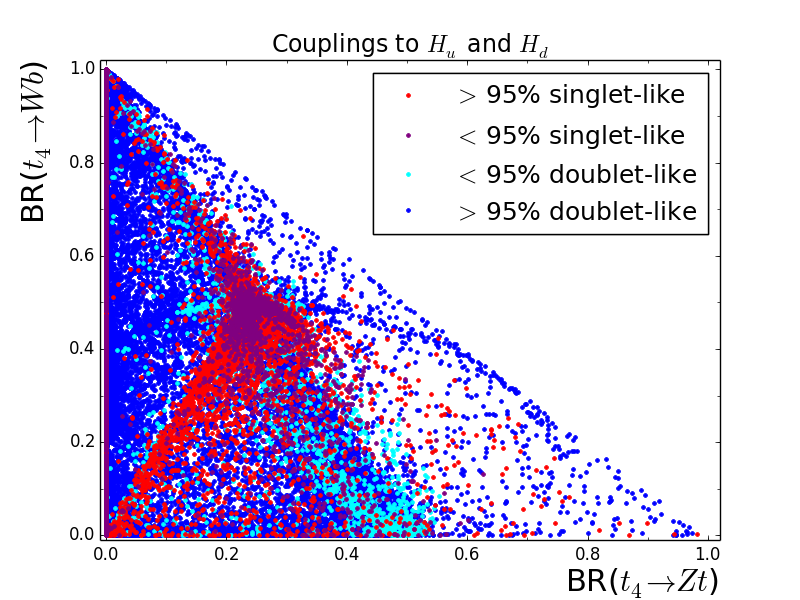}
\includegraphics[width=.49\linewidth]{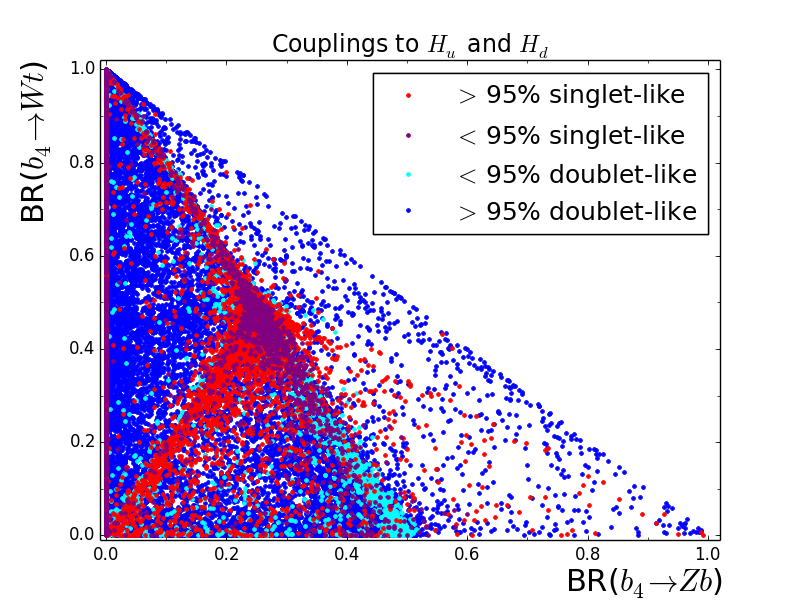} 
\caption{
The allowed branching ratios of the lightest vectorlike quarks in the scenario with couplings to $H_u$ only (top-left), couplings to $H_d$ only (top-right), and the general scenario (bottom panels). Red: 95\% or more singlet-like, purple: 50\%-95\% singlet-like, cyan: 50\%-95\%  doublet-like, blue: 95\% or more doublet-like.
}
\label{fig:BRscenarios}
\end{center}
\end{figure}
\begin{figure}
\begin{center}
\includegraphics[width=.49\linewidth]{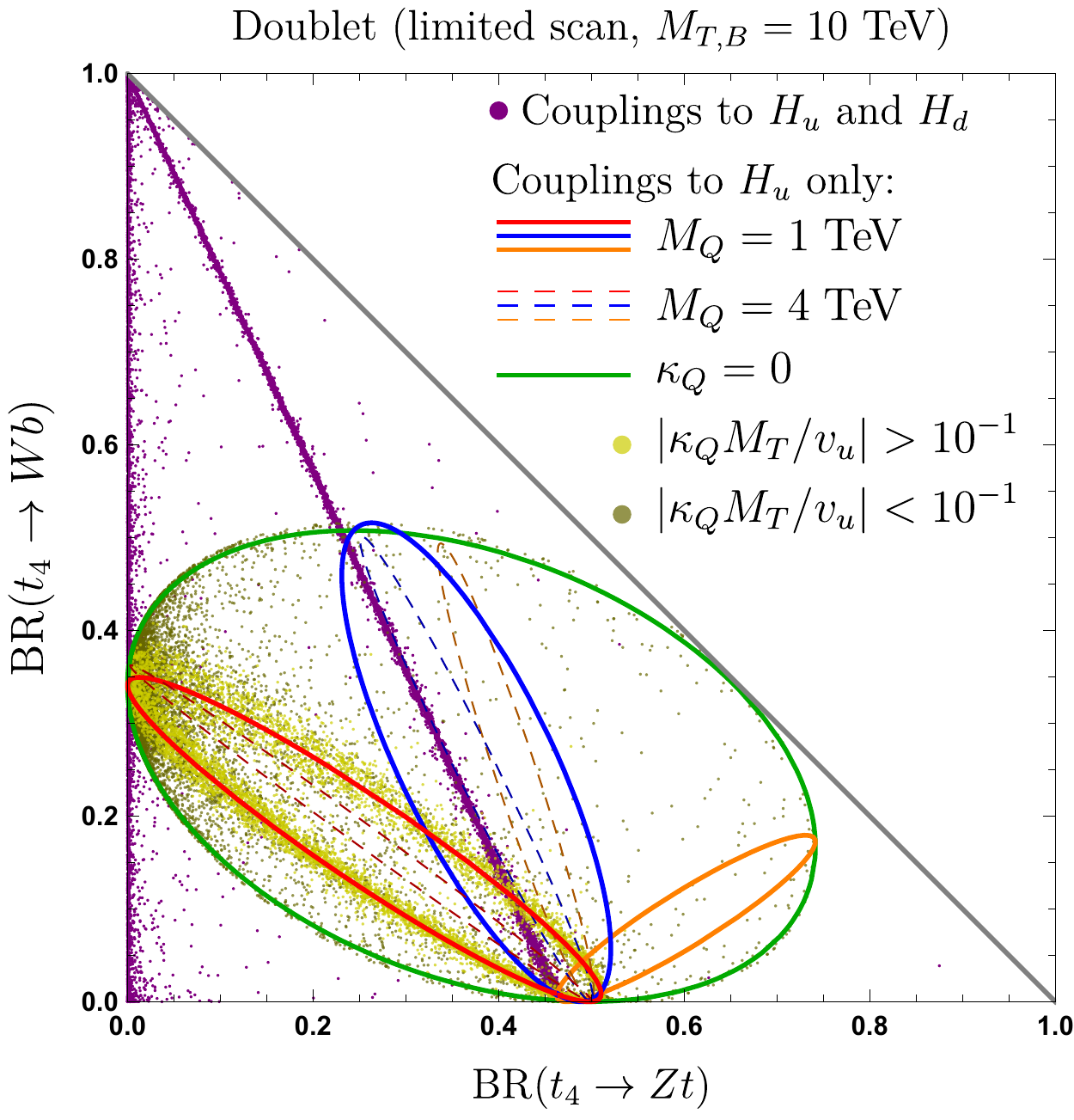} 
\includegraphics[width=.49\linewidth]{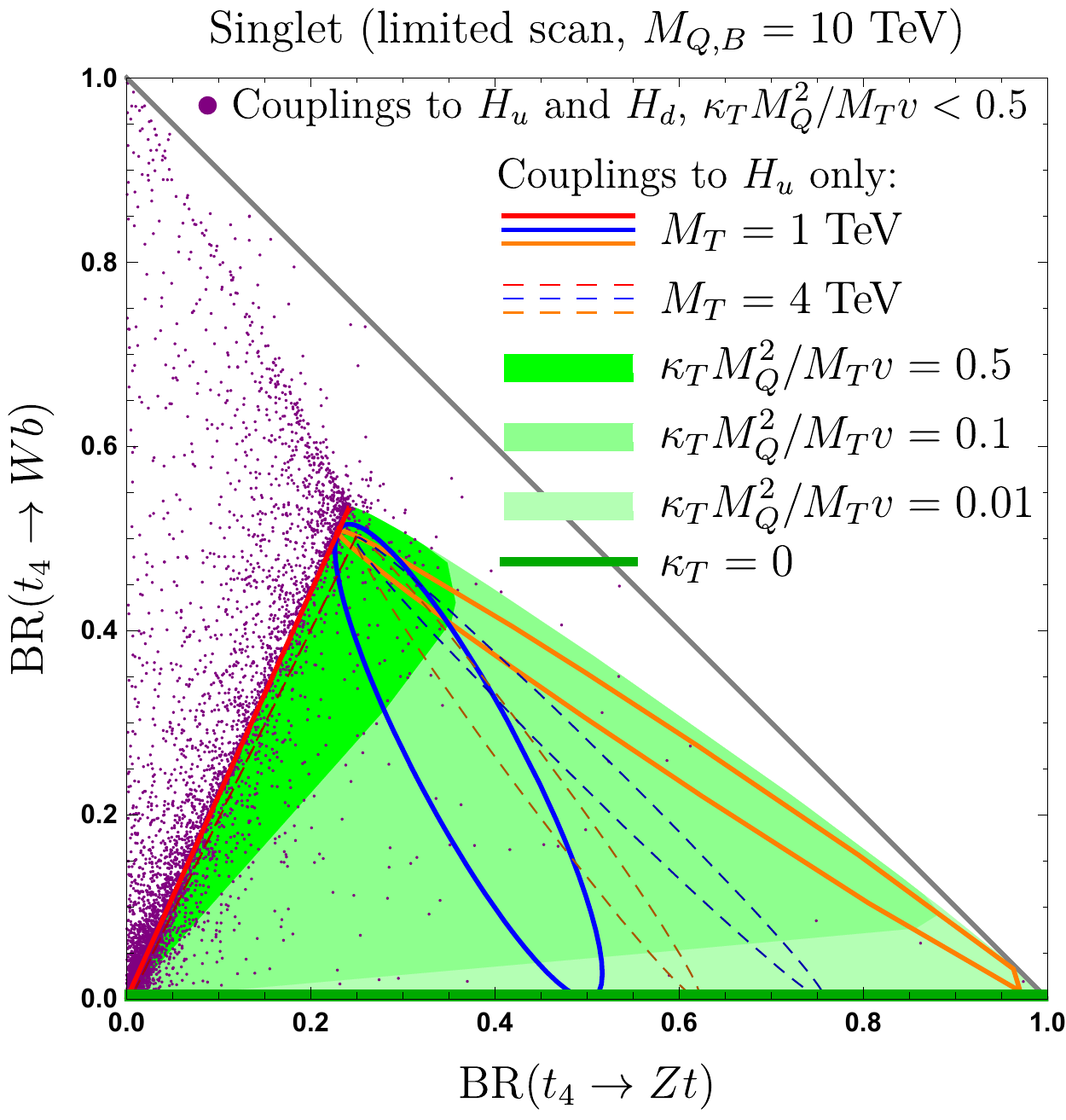} 
\caption{The left (right) panel shows the allowed branching ratios of a mostly doublet (singlet) $t_4$. The green ellipse (line) is obtained for $\kappa_Q=0$ ($\kappa_T=0$). The red, blue and orange ellipses show three cases of how a $\kappa_Q=0$ ($\kappa_T=0$) point moves towards the Goldstone boson equivalence limit as $\kappa_Q$ ($\kappa_T$) is increased. Solid and dashed lines correspond to $M_T = (1,4)\; {\rm TeV}$ ($M_Q = (1,4)\; {\rm TeV}$). Purple points correspond to a coarse scan of the whole parameter space. For the doublet case, yellow (green) points correspond to $|\kappa_Q M_T/v_u| > 10^{-1}$ ($|\kappa_Q M_T/v_u| < 10^{-1}$); for the singlet case, the shaded green regions show the effect of decreasing $\kappa_T M_Q^2/(M_T v)$. 
}
\label{fig:BRscenarios_cartoon}
\end{center}
\end{figure}

The results of the parameter space scan for decays into $W$, $Z$ and $h$ are summarized in Fig.~\ref{fig:BRscenarios}, where we show branching ratios of the lightest up-type and down-type vectorlike quarks $t_4$ and $b_4$. In the top two panels we consider the limiting cases of couplings to $H_u$ and $H_d$ only for which we consider only the vectrolike quark with non-trivial branching ratios; in fact, in presence of couplings to $H_u$ ($H_d$) only, the lightest vectorlike quark $b_4$ ($t_4$) decays exclusively to $W(t,t_4)$ ($W(b,b_4)$).\footnote{Note that, in presence of couplings to $H_u$ ($H_d$) only, the decay mode $t_4 \to b_4 W$ ($b_4\to t_4 W$) is not allowed. Considering the $t_4$ case (the $b_4$ one is completely analogue) this fact can be understood as follows. If the $b_4$ is a singlet, the absence of mixing in the down sector causes the $Wt_4 b_4$ vertex to vanish. If the $b_4$ is a doublet, in absence of mixing its mass is exactly $M_Q$. The mass matrix for $t_4$ and $t_5$ has $M_Q$ and $M_T$ on the diagonal with some small mixing, implying that $m_{t_4} < \text{min}(M_Q,M_T)$. In both cases the decay $t_4 \to b_4 W$ is kinematically not allowed: for doublet $t_4$ ($M_Q<M_T$) we have $m_{t_4} < M_Q = m_{b_4}$ and for singlet $t_4$ ($M_T<M_Q$) we have $m_{t_4} < M_T <  M_Q = m_{b_4}$.} 

In the bottom panels we consider the general case. In presence of couplings to both $H_u$ and $H_d$ the decays $t_4 \to b_4 W$ and $b_4 \to t_4 W$ can be present. In this case, we rescale the branching ratios by the sums, ${\rm BR}(t_4 \to h t) + {\rm BR}(t_4 \to W b) + {\rm BR}(t_4 \to Z t)$ and ${\rm BR}(b_4 \to h b) + {\rm BR}(b_4 \to W t) + {\rm BR}(b_4 \to Z b)$, respectively. The rescaled branching ratios into $h$, $W$ and $Z$ shown in Fig.~\ref{fig:BRscenarios} add to one (the values read from the figure are the upper limits for the actual branching ratios).
Red, purple, cyan, and blue points correspond to $d_{t_4,b_4}<0.05$ ($s_{t_4,b_4}>0.95$), $0.05<d_{t_4,b_4}<0.50$ ($0.50<s_{t_4,b_4}<0.95$), $0.5<d_{t_4,b_4}<0.95$ and $d_{t_4,b_4}>0.95$, respectively.  
Note that if the decay modes into heavy Higgses are kinematically open all plots in Fig.~\ref{fig:BRscenarios} correspond to rescaled branching ratios. 

The main features of these plots can be understood using the approximate formulas presented in eqs.~(\ref{eq:doubletlam34})-(\ref{eq:singletgRZ43}), which we use to generate the two plots in figure~\ref{fig:BRscenarios_cartoon}. The following very simplified expressions for the $h$, $W$ and $Z$ couplings for a mostly doublet $t_4$ yield the correct branching ratios up to terms suppressed by $(\kappa's,\lambda's) v/M_{Q,T,B}$:
\begin{align}
\lambda^h_{t t_4} & \simeq -2 \frac{v_u}{M_T} \kappa_T \bar \kappa \sin\beta \, , \label{eq:simpl1}\\
\lambda^h_{t_4 t} & \simeq \kappa_Q \sin\beta \, ,\\
g_L^{Wt_4 b} & \simeq -\frac{g}{\sqrt{2}} \frac{v}{M_Q} \left[
\frac{v_u}{M_T} \kappa_T \bar \kappa \sin\beta  -  \frac{v_d}{M_B} \lambda_B \bar \lambda \cot\beta \right]  \, ,\label{eq:simpl3}\\
g_R^{Wt_4 b} & \simeq -\frac{g}{\sqrt{2}} \frac{v}{M_Q} \left[ -\lambda_Q \cos\beta\right] \, ,\\
g_L^{Zt_4 t} & \simeq 0 \, ,\\
g_R^{Zt_4 t} & \simeq  -\frac{g}{2\cos\theta_W} \frac{v}{M_Q} \left[ \kappa_Q \sin\beta
  \right] \, .\label{eq:simpl2}
\end{align}
Note that the $v/M_Q$ factor in the $W$ and $Z$ couplings is compensated by the $M_Q^2 /M_{W,Z}^2$ factors that appear in the formulae for the decay widths, see the discussion below \eqref{eq:singletgRZ43}.

We start by considering the case of a mostly doublet $t_4$ with couplings to $H_u$ only (i.e. all $\lambda$'s set to zero). In the Goldstone boson equivalence limit we see that only couplings to $Z$ and $h$ are non-zero and that they are controlled by the parameter $\kappa_Q$, implying ${\rm BR}(t_4\to Zt) = {\rm BR}(t_4\to ht) =1/2$, that corresponds to the point $[1/2,0]$ in the $[{\rm BR}(t_4\to Zt),{\rm BR}(t_4\to W b)]$ plane. Corrections to this limiting case are suppressed by at least one mixing parameter ($\kappa$, $\bar \kappa$) and by additional powers of the heavy singlet mass $M_T$ (which for mostly doublet $t_4$ is larger than $M_Q$). 

Looking at the subleading contributions in Eqs.~(\ref{eq:simpl1})-(\ref{eq:simpl2}), we see that the widths into $h$, $W$ and $Z$ are controlled by two combinations of parameters ($\kappa_Q \sin\beta$ and $\frac{v_u}{M_T} \kappa_T \bar \kappa \sin\beta$), implying that as $\kappa_Q$ decreases while all other $\kappa$'s remain sizable, the branching ratios move into an ellipse in the $[{\rm BR}(t_4\to Zt),{\rm BR}(t_4\to W b)]$ plane (red in the left panel of figure~\ref{fig:BRscenarios_cartoon}) as discussed in appendix~\ref{ellipses}. 
For $\kappa_Q= 0$ the dominant couplings are $\lambda^h_{t t_4}$ and $g_L^{Wt_4 t} =\frac{g}{\sqrt{2}} \frac{v}{M_Q} \frac{\lambda^h_{t t_4}}{2}$, which imply ${\rm BR}(t_4\to Wb)=1/3$ and ${\rm BR}(t_4\to ht)=2/3$ (See the note at the end of appendix~\ref{app:special}), corresponding to the point $[0,1/3]$ in the figure (the end point of the red ellipse). 

Including all the subleading dependence on the $\kappa$'s (i.e. terms suppressed by at least two  powers of the heavy vectorlike mass $M_T$) one obtains the formulas in eqs.~(\ref{eq:doubletlam34})-(\ref{eq:singletgRZ43}). For $\kappa_Q$ similar in size to $(v_u/M_T) \kappa \bar \kappa$ a scan deviates mildly from the red ellipse (yellow points); if $\kappa_Q$ is allowed to be very small, the subleading terms mentioned above (green points) fill completely the green ellipse. For $\kappa_Q =0$ the six couplings depend on the two combinations $(v_u/M_T) \kappa_T \bar \kappa \sin\beta$ and $(v_u M_Q/M_T^2) \kappa_T \kappa \sin\beta$, implying that the branching ratios lie exactly on the green ellipse.\footnote{Starting from a point on the ellipse one quickly moves towards the GBEL point $[1/2,0]$ as $\kappa_Q$ increases. The resulting path in the plane for three representative points is described by the red, blue and orange ellipses. The minor axes of the ellipses decreases for heavier vectorlike quark masses. Since the points on the right side of ellipses are considerably fine-tuned, as soon as we change the vectorlike quark mass (without changing the other parameters), the points move towards the left-hand side (as is the case for the orange curves). } In particular, note that unless $\bar \kappa \ll \kappa M_Q/M_T$ only the region of the ellipse near the point $(0,1/3)$ is allowed; this is reflected in the high density of points near this limit at small $\kappa_Q$. This feature can be seen also in the full scan presented in the upper panels of figure~\ref{fig:BRscenarios}, in which the right-hand side of the ellipse is sparsely populated.

Finally, allowing also couplings to $H_d$ all branching ratios on the segment between $[1/2,0]$ and $[0,1]$ are allowed. In fact, branching ratios into $h$ and $Z$ are controlled by $\kappa_Q$ (and are identical in the limit in which all other $\kappa$'s are neglected) while the branching ratio into $W$ is controlled by $\lambda_Q$. A limited parameter space scan yields points (purple) clustered around this line. If only $\kappa_Q (\lambda_Q)$ is present the points cluster around $[1/2,0]$ ($[0,1]$). On the other hand, if $\kappa_Q = \lambda_Q = 0$, there are only sizable contributions to $\lambda^h_{tt_4}$ and $g_L^{Wt_4 b}$ (see eqs.~(\ref{eq:simpl1}) and (\ref{eq:simpl3})) which are now controlled by two independent combinations of parameters, implying that decay into $Z$ is suppressed and the remaining two channels can have any branching ratios. This can be seen by the purple points clustered on the vertical line corresponding to ${\rm BR} (t_4\to Z t)=0$ in figure~\ref{fig:BRscenarios_cartoon}. In the high-density scan presented in the bottom two panels of figure~\ref{fig:BRscenarios} we see that every branching ratio is  possible, but that the region in between the two purple lines in figure~\ref{fig:BRscenarios_cartoon} is significantly more populated. Note that, if there are couplings only to $H_d$, $t_4$ decays exclusively into the channels $Wb$ and $Wb_4$.

In the case of a mostly singlet $t_4$ with couplings to $H_u$ only (i.e. all $\lambda$'s set to zero) the situation is very different. Keeping only the leading terms in $1/M_{Q,T,B}$ for each coupling we obtain the following very simplified expressions for the $h$, $W$ and $Z$ couplings for a mostly singlet $t_4$: 
\begin{align}
\lambda^h_{t t_4} & \simeq 
\kappa_T \sin\beta 
\, , \label{eq:ssimpl1}\\
\lambda^h_{t_4 t} & \simeq 
 -2 \sin\beta \bar\kappa \kappa_Q \frac{v_u}{M_Q} \, ,\\
g_L^{Wt_4 b} & \simeq \frac{g}{\sqrt{2}} \frac{v}{M_T} \left[
\kappa_T \sin\beta\right]
  \, ,\label{eq:ssimpl3}\\
g_R^{Wt_4 b} & \simeq \frac{g}{\sqrt{2}} \frac{v}{M_T} \left[
\kappa \frac{v_d M_T}{M_Q^2}\lambda_Q \sin\beta\right] \, ,\label{eq:ssimpl4}\\
g_L^{Zt_4 t} & \simeq \frac{g}{2\cos\theta_W} \frac{v}{M_T} \left[\kappa_T \sin\beta \right]\, ,\\
g_R^{Zt_4 t} & \simeq  \frac{g}{2\cos\theta_W} \frac{v}{M_T} \left[ \kappa\kappa_Q \sin\beta \frac{v_u M_T}{M_Q^2}
  \right] \, .\label{eq:ssimpl2}
\end{align}
In the GBEL the $h$, $W$ and $Z$ couplings are controlled by the parameter $\kappa_T$, implying ${\rm BR}(t_4\to ht) = {\rm BR}(t_4\to Zt) = \frac{1}{2} {\rm BR}(t_4\to Wb) =1/4$,
that corresponds to the point $[1/4,1/2]$ in the $[{\rm BR}(t_4\to Zt),{\rm BR}(t_4\to W b)]$ plane. Corrections to this limiting case are suppressed by at least one mixing parameter ($\kappa$, $\bar \kappa$) and by additional powers of the heavy doublet mass $M_Q$ (which for mostly singlet $t_4$ is much larger than $M_T$). 

As $\kappa_T$ decreases below $v/M_Q$, the largest coupling becomes  $\lambda^h_{t_4 t}$, implying that the branching ratio into $h$ increases and we move along a line towards the origin in the $[{\rm BR}(t_4\to Zt),{\rm BR}(t_4\to W b)]$ plane (see the red line in the right panel of figure~\ref{fig:BRscenarios_cartoon}). As $\kappa_T$ further decreases below $v M_T/M_Q^2$, the right-handed $Z$ coupling becomes sizable and a large $t_4 \to Z t$ branching ratio becomes possible (see green shaded triangle in the right panel of figure~\ref{fig:BRscenarios_cartoon}). 
For $\kappa_T = 0$, the $W$ channel disappears and the points collapse on the line at ${\rm BR}(t_4\to Wb) = 0$ (green line in the right panel of figure~\ref{fig:BRscenarios_cartoon}); note that points with large ${\rm BR}(t_4\to Zt)$ require some degree of fine tuning.\footnote{Starting from a point on the $\kappa_T=0$ line one quickly moves towards the GBEL point $[1/4,1/2]$ as $\kappa_T$ increases. The resulting path in the plane for three representative points is described by the red, blue and orange ellipses. The minor axes of the ellipses decreases for heavier vectorlike quark masses. Since the points at large ${\rm BR} (t_4\to Z t)$ are considerably fine-tuned, as soon as we change the vectorlike quark mass (without changing the other parameters), the points move considerably, as is the case for the orange and blue curves.}

Finally, allowing also couplings to $H_d$, we see that the $t_4\to Wb$ branching ratio becomes effectively independent of the others for $\kappa_T < v M_T/M_Q^2$ as we see from eq.~(\ref{eq:ssimpl4}). Under these conditions we start moving along a line towards the point [0,1] in the $[{\rm BR}(t_4\to Zt),{\rm BR}(t_4\to W b)]$ plane (see purple points in the right panel of figure~\ref{fig:BRscenarios_cartoon}). 

The features described here help  understanding the results of the full scan presented in the figure~\ref{fig:BRscenarios}. Note that the $b_4$ case is almost identical to the $t_4$ one, the only difference being the replacement of the top Yukawa and mass with the bottom ones.

\section{Conclusions}
\label{sec:conclusions}

We have analyzed decays of new quarks  in extensions of the standard model and a two Higgs doublet model (type-II) by vectorlike pairs of new quarks, corresponding to a copy of  SM quark SU(2) doublets and singlets and their vectorlike partners. We assumed only mixing with the third generation of SM quarks, nevertheless the results can be straightforwardly generalized for cases of mixing with the first or second generation. 

We have identified several typical  patterns of branching ratios of the lightest new up-type quark, $t_4$, and down-type quark, $b_4$, depending on the structure of Yukawa couplings that mix the vectorlike and standard model quarks  and also on their doublet or singlet nature. Among the most striking results is the finding that decays  into heavy neutral or charged Higgs bosons, when kinematically open,  can easily dominate and even be close to 100\%:  $b_4\to H b$ at medium to large $\tan \beta$, $t_4\to H t$ at small  $\tan \beta$ and $b_4\to H^\pm t$, $t_4\to H^\pm b$ at both large and small $\tan \beta$.

We found that the conventional decay modes into $W$, $Z$ and the SM Higgs boson, $h$, follow the pattern expected from the Goldstone boson equivalence limit that we have generalized to scenarios with all possible couplings. For doublet-like new quarks this leads to a one parameter family of branching ratios characterized by an arbitrary branching ratio to $W$  and equal branching ratios to $Z$ and $h$. 

We have also discussed in very detail the structure of Yukawa couplings required to significantly deviate from the pattern characteristic of Goldstone boson equivalence limit  that can result in essentially arbitrary branching ratios. For vectorlike quark masses within the reach of the LHC and near future colliders this does not require very special choices of model parameters. We found that large deviations from the GBEL are possible for $(\kappa_Q,\lambda_Q) < v/M_T$ and $\kappa_T < v M_T/M_Q^2$ for doublet-like and singlet-like $t_4$, respectively (similar relations hold for $b_4$ with $M_T\to M_B$ and $\kappa_T\to \lambda_B$). 

The new decay modes of vectorlike quarks through heavy Higgs bosons and the fact that they easily dominate imply that the usual search strategies are not sufficient and the current exclusion limits might not necessarily apply. Among the smoking gun signatures are the $6 t$, $4t 2b$, $2t4b$ and $6b$ final states resulting from the pair production of vectorlike quarks  (and  similar final states for single production). The SM backgrounds for  these final states (at large invariant mass)  are very small and thus searching for these processes could lead to the simultaneous discovery of a new Higgs boson and a new quark. 

\acknowledgments 
SS thanks Bogdan Dobrescu and Yuval Grossman for insightful discussion. We thank Navin McGinnis for invaluable help in cross checking the analytic results. The work of RD was supported in part by the U.S. Department of Energy under grant number {DE}-SC0010120. This work was performed in part at the Aspen Center for Physics, which is supported by National Science Foundation grant PHY-1607611.
SS appreciates the hospitality of Fermi National Accelerator Laboratory. 
SS is supported by the National Research Foundation of Korea (NRF-2017R1D1A1B03032076) and in partial by the NRF grant funded by the Korean government (MISP) (No.2016R1A2B2016112).

\appendix

\section{Details of the Model}
\label{app:model}
\subsection{Mass eigenstates}
From the Lagrangian in eq.~(\ref{eq:lagrangian}) we read the $5\times 5$ mass matrices in the up and down sectors in the gauge eigenstate basis:
\begin{align}
\left( \begin{array}{ccc}\bar u_{Li} & \bar T_L^Q & \bar T_L\end{array}\right)
\left( \begin{array}{ccc}
y_u^{ij} v_u & 0 & \kappa_T^i v_u \\
\kappa_Q^j v_u & M_Q & \kappa v_u \\
0 & \bar \kappa v_u & M_T \\
\end{array}\right)
\left(\begin{array}{c}u_{Rj}  \\T_R^Q \\T_R\end{array}\right)~,
\label{eq:mmu5} \\
\left( \begin{array}{ccc}\bar d_{Li} & \bar B_L^Q & \bar B_L\end{array}\right)
\left( \begin{array}{ccc}
y_d^{ij} v_d & 0 & \lambda_B^i v_d \\
\lambda_Q^j v_d & M_Q & \lambda v_d \\
0 & \bar \lambda v_d & M_B \\
\end{array}\right)
\left(\begin{array}{c}d_{Rj}  \\B_R^Q \\B_R\end{array}\right)~,
\label{eq:mmd5}
\end{align}
where $i, j = 1, 2, 3$. It is convenient to define the following vectors of flavor eigenstates: $u_{La} = ( u_{L1}, u_{L2}, u_{L3}, T^Q_L, T_L)$ and $d_{La} = ( d_{L1}, d_{L2}, d_{L3}, B^Q_L, B_L)$, and similarly for $u_{Ra}$ and $d_{Ra}$. 

We then use four unitary matrices to diagonalize the Yukawa matrices $y_u^{ij}$ and $y_d^{ij}$:
\begin{align}
y_u^{\rm diag} &= U_L^\dagger y_u U_R \; , \label{yudiag}\\
y_d^{\rm diag} &= D_L^\dagger y_u D_R \; . \label{yddiag}
\end{align}
The resulting up and down mass matrices are:
\begin{align}
\left( \begin{array}{ccc}
y_u^{\rm diag} v_u & 0 & (U^\dagger_L)^{il}\kappa_T^l v_u \\
\kappa_Q^l (U_R)^{lj} v_u & M_Q & \kappa v_u \\
0 & \bar \kappa v_u & M_T \\
\end{array}\right)
\longrightarrow \left(
\begin{array}{ccccc}
m_u & 0 & 0 & 0 & 0 \\
0 & m_c & 0 & 0 & 0 \\
0 & 0 & y_t v_u & 0 & \kappa_T v_u \\
0 & 0 & \kappa_Q v_u & M_Q & \kappa v_u \\
0 & 0 & 0 & \bar \kappa v_u & M_T 
\end{array}
\right) \equiv
\left(
\begin{array}{ccc}
m_u & 0 & 0 \\
0 & m_c & 0 \\
0 & 0 & M_t 
\end{array}
\right) 
\; , \label{eq:mass1}\\
\left( \begin{array}{ccc}
y_d^{\rm diag} v_d & 0 & (D^\dagger_L)^{il}\lambda_B^l v_d \\
\lambda_Q^l (D_R)^{lj} v_d & M_Q & \lambda v_d \\
0 & \bar \lambda v_d & M_B \\
\end{array}\right)
\longrightarrow 
\left(
\begin{array}{ccccc}
m_d & 0 & 0 & 0 & 0 \\
0 & m_s & 0 & 0 & 0 \\
0 & 0 & y_b v_d & 0 & \lambda_B v_d \\
0 & 0 & \lambda_Q v_d & M_Q & \lambda v_d \\
0 & 0 & 0 & \bar \lambda v_d & M_B 
\end{array}
\right)\equiv
\left(
\begin{array}{ccc}
m_d & 0 & 0 \\
0 & m_s & 0 \\
0 & 0 & M_b 
\end{array}
\right)  \; ,
\label{eq:mass2}
\end{align}
where the requirement of mixing with the third generation only has been imposed by setting $(U^\dagger_L)^{il}\kappa_T^l = \kappa_T \delta_{i3}$, $\kappa_Q^l (U_R)^{lj} = \kappa_Q \delta_{j3}$, $(D^\dagger_L)^{il}\lambda_B^l = \lambda_B \delta_{i3}$ and $\lambda_Q^l (D_R)^{lj} = \lambda_Q \delta_{j3}$. The basis vectors corresponding to the lower-right $3\times 3$ block are $(t_L, T^Q_L, T_L)$ and $(b_L, B^Q_L, B_L)$ (and similarly for right-handed components), see eqs.~(\ref{eq:mmu}) and (\ref{eq:mmd}).

Finally we diagonalize the $3\times 3$ mass matrices $M_t$ and $M_b$:
\begin{align}
V_L^{u\, \dagger} M_t V_R^u & = 
\left(
\begin{array}{c c c}
m_t & 0 & 0 \\
0 & m_{t_4} & 0 \\
0 & 0 & m_{t_5}
\end{array}
\right) \; , \\
V_L^{d\, \dagger} M_b V_R^d & = 
\left(
\begin{array}{c c c}
m_b & 0 & 0 \\
0 & m_{b_4} & 0 \\
0 & 0 & m_{b_5}
\end{array}
\right) \; .
\end{align}
The vectors ${\hat u}_{La}$ and ${\hat d}_{La}$ of mass eigenstates (and similarly for right-handed fields) are explicitly given by: 
\begin{align}
{\hat u}_{La} 
&= 
\left(\begin{array}{c} 
{\hat u}_{Li} \cr {\hat t}_{L4} \cr {\hat t}_{L5} \cr 
\end{array}\right)
= 
\left(\begin{array}{cc} 
1_{2\times 2} & 0_{2\times 3} \cr
0_{3\times 2} & V^{u\dagger}_L\cr
\end{array}\right)
\left( \begin{array}{ccc} 
 U_L^\dagger &  0_{3\times 1} & 0_{3\times 1}  \cr
 0_{1\times 3}  & 1 & 0 \cr
 0_{1\times 3} & 0 & 1 \cr
\end{array}\right)
\left(\begin{array}{c} 
 u_{Li} \cr T^Q_L \cr T_L \cr 
\end{array}\right) \; ,
\\
{\hat d}_{La} 
&= 
\left(\begin{array}{c} 
{\hat d}_{Li} \cr {\hat b}_{L4} \cr {\hat b}_{L5} \cr 
\end{array}\right)
= 
\left(\begin{array}{cc} 
1_{2\times 2} & 0_{2\times 3} \cr
0_{3\times 2} & V^{d\dagger}_L\cr
\end{array}\right)
\left( \begin{array}{ccc} 
 D_L^\dagger &  0_{3\times 1} & 0_{3\times 1}  \cr
 0_{1\times 3}  & 1 & 0 \cr
 0_{1\times 3} & 0 & 1 \cr
\end{array}\right)
\left(\begin{array}{c} 
 d_{Li} \cr B^Q_L \cr B_L \cr 
\end{array}\right) \; . 
\end{align}

It is useful to have approximate expressions in the limit $\kappa_T v_u$, $\kappa_Q v_u$, $\kappa v_u$, $\bar \kappa v_u \ll M_Q$, $M_T$, $|M_Q - M_T|$ and $\lambda_B v_d$, $\lambda_Q v_d$, $\lambda v_d$, $\bar \lambda v_d \ll M_Q$, $M_B$, $|M_Q - M_B|$. If the lightest vectorlike quarks ($t_4$, $b_4$) are mostly $SU(2)$ doublets, we find:
\dis{
V_L^u &= 
\left(
\begin{array}{c c c}
1  - v_u^2 \, \frac{\kappa_T^2}{2M_T^2} & -v_u^2 \left(\frac{\kappa_T}{M_Q} \frac{\kappa M_Q + \bar \kappa M_T}{M_T^2 - M_Q^2} -\frac{y_t \kappa_Q}{M_Q^2} \right) &  v_u \frac{\kappa_T}{M_T} \\
v_u^2 \, \frac{\kappa_T \bar \kappa M_Q - y_t \kappa_Q M_T}{M_Q^2 M_T} & ~~1- v_u^2 \, \frac{(M_Q \bar \kappa + M_T \kappa)^2}{2(M_T^2 - M_Q^2)^2} & v_u \frac{(M_Q \bar \kappa + M_T \kappa)}{M_T^2 - M_Q^2} \\
- v_u \, \frac{\kappa_T}{M_T} & - v_u \frac{(M_Q \bar \kappa + M_T \kappa)}{M_T^2 - M_Q^2} &  ~~1- v_u^2 \, \frac{\kappa_T^2}{2M_T^2} - v_u^2 \, \frac{(M_Q \bar \kappa + M_T \kappa)^2}{2(M_T^2 - M_Q^2)^2} 
\end{array}
\right)~,
\label{eq:VLu}
}
\dis{
V_R^u &=
\left(
\begin{array}{ccc}
1 - v_u^2 \, \frac{\kappa_Q^2}{2 M_Q^2} & v_u \, \frac{\kappa_Q}{M_Q} & v_u^2 \left( \frac{\kappa_Q}{M_T} \frac{M_Q \bar \kappa + M_T \kappa}{M_T^2 - M_Q^2} + \frac{y_t \kappa_T}{M_T^2} \right) \\
-v_u \, \frac{\kappa_Q}{M_Q} & ~~1 - v_u^2 \, \frac{\kappa_Q^2}{2 M_Q^2} - v_u^2 \, \frac{(M_Q \kappa + M_T \bar \kappa)^2}{2(M_T^2  - M_Q^2)^2} & v_u \frac{(M_Q \kappa + M_T \bar \kappa)}{M_T^2  - M_Q^2} \\
v_u^2 \, \frac{\kappa_Q \bar \kappa M_T - y_t \kappa_T M_Q}{M_Q M_T^2} & - v_u \frac{(M_Q \kappa + M_T \bar \kappa)}{M_T^2  - M_Q^2} & ~~1 - v_u^2 \, \frac{(M_Q \kappa + M_T \bar \kappa)^2}{2(M_T^2  - M_Q^2)^2} \\
\end{array}
\right)~,
\label{eq:VRu}
}
\dis{
V_L^d &= 
\left(
\begin{array}{c c c}
1  - v_d^2 \, \frac{\lambda_B^2}{2M_B^2} & -v_d^2 \left(\frac{\lambda_B}{M_Q} \frac{\lambda M_Q + \bar \lambda M_B}{M_B^2 - M_Q^2} -\frac{y_b \lambda_Q}{M_Q^2} \right) &  v_d \frac{\lambda_B}{M_B} \\
v_d^2 \, \frac{\lambda_B \bar \lambda M_Q - y_b \lambda_Q M_B}{M_Q^2 M_B} & ~~1- v_d^2 \, \frac{(M_Q \bar \lambda + M_B \lambda)^2}{2(M_B^2 - M_Q^2)^2} & v_d \frac{(M_Q \bar \lambda + M_B \lambda)}{M_B^2 - M_Q^2} \\
- v_d \, \frac{\lambda_B}{M_B} & - v_d \frac{(M_Q \bar \lambda + M_B \lambda)}{M_B^2 - M_Q^2} &  ~~1- v_d^2 \, \frac{\lambda_B^2}{2M_B^2} - v_d^2 \, \frac{(M_Q \bar \lambda + M_B \lambda)^2}{2(M_B^2 - M_Q^2)^2} 
\end{array}
\right)~,
\label{eq:VLd}
}
\dis{
V_R^d &=
\left(
\begin{array}{ccc}
1 - v_d^2 \, \frac{\lambda_Q^2}{2 M_Q^2} & v_d \, \frac{\lambda_Q}{M_Q} & v_d^2 \left( \frac{\lambda_Q}{M_B} \frac{M_Q \bar \lambda + M_B \lambda}{M_B^2 - M_Q^2} + \frac{y_b \lambda_B}{M_B^2} \right) \\
-v_d \, \frac{\lambda_Q}{M_Q} & ~~1 - v_d^2 \, \frac{\lambda_Q^2}{2 M_Q^2} - v_d^2 \, \frac{(M_Q \lambda + M_B \bar \lambda)^2}{2(M_B^2  - M_Q^2)^2} & v_d \frac{(M_Q \lambda + M_B \bar \lambda)}{M_B^2  - M_Q^2} \\
v_d^2 \, \frac{\lambda_Q \bar \lambda M_B - y_b \lambda_B M_Q}{M_Q M_B^2} & - v_d \frac{(M_Q \lambda + M_B \bar \lambda)}{M_B^2  - M_Q^2} & ~~1 - v_d^2 \, \frac{(M_Q \lambda + M_B \bar \lambda)^2}{2(M_B^2  - M_Q^2)^2} \\
\end{array}
\right)~,
\label{eq:VRd}
}
up to corrections of $\mathcal{O}(\epsilon^3)$ where $\epsilon = (\kappa_Q, \kappa_T, \kappa, \bar \kappa) v_u /(M_Q,M_T)$ or $(\lambda_Q, \lambda_B, \lambda, \bar \lambda) v_d /(M_Q,M_B)$.  The corresponding formulas for the lightest vectorlike quarks being mostly $SU(2)$ singlets can be obtained by swapping the second and third column in each mixing matrix. Note that in the numerical analysis we use the exact expressions.

\subsection{Couplings of the W boson}
\label{sec:wboson}
The Lagrangian in eq.~(\ref{eq:lagrangian}) yields the following $W$ boson interactions for left-handed and right-handed fermions:
\begin{align}
{\cal L}_W^L &= 
\frac{g}{\sqrt{2}}
\left( \begin{array}{ccccc} \bar u_{Li}   \cr \bar T_L^Q  \cr \bar T_L \end{array}\right)^T 
\slashed W
\left( \begin{array}{ccc} 
 1_{3\times 3} &  0_{3\times 1} & 0_{3\times 1}  \cr
 0_{1\times 3}  & 1 & 0 \cr
 0_{1\times 3} & 0 & 0 \cr
\end{array}\right)
\left( \begin{array}{c}  d_{Li} \cr  B_L^Q \cr  B_L \end{array}\right)
+ \text{h.c.}\\
&= 
\frac{g}{\sqrt{2}}
\left( \begin{array}{ccccc} \bar {\hat u}_{Li} \cr \bar {\hat t}_{L4} \cr \bar {\hat t}_{L5} \end{array}\right)^T
\slashed W
\left( \begin{array}{cc} 
1_{2\times 2} & 0_{2\times 3} \cr
0_{3\times 2} & (V^u_L)^\dagger\cr
\end{array}\right)
\left( \begin{array}{ccc} 
 \hat V_{\rm CKM} &  0_{3\times 1} & 0_{3\times 1}  \cr
 0_{1\times 3}  & 1 & 0 \cr
 0_{1\times 3} & 0 & 0 \cr
\end{array}\right)
\left( \begin{array}{cc} 
1_{2\times 2} & 0_{2\times 3} \cr
0_{3\times 2} & V^d_L\cr
\end{array}\right)
\left( \begin{array}{c}  \hat d_{Li} \cr   {\hat b}_{L4} \cr  { \hat b}_{L5} \end{array}\right) + \text{h.c.} \label{eq:WL}\\
{\cal L}_W^R &= 
\frac{g}{\sqrt{2}}
\left( \begin{array}{ccccc} \bar {\hat u}_{Ri}  \cr \bar T_R^Q  \cr \bar T_R \end{array}\right)^T 
\slashed W
\left( \begin{array}{ccc} 
 0_{3\times 3} &  0_{3\times 1} & 0_{3\times 1}  \cr
 0_{1\times 3}  & 1 & 0 \cr
 0_{1\times 3} & 0 & 0 \cr
\end{array}\right)
\left( \begin{array}{c}  d_{Ri} \cr  B_R^Q \cr  B_R \end{array}\right)
+ \text{h.c.}\\
&= 
\frac{g}{\sqrt{2}}
\left( \begin{array}{ccccc} \bar {\hat u}_{Ri}  \cr \bar {\hat t}_{R4} \cr \bar {\hat t}_{R5} \end{array}\right)^T
\slashed W
\left( \begin{array}{cc} 
1_{2\times 2} & 0_{2\times 3} \cr
0_{3\times 2} & (V^u_R)^\dagger\cr
\end{array}\right)
\left( \begin{array}{ccc} 
  0_{3\times 3} &  0_{3\times 1} & 0_{3\times 1}  \cr
 0_{1\times 3}  & 1 & 0 \cr
 0_{1\times 3} & 0 & 0 \cr
\end{array}\right)
\left( \begin{array}{cc} 
1_{2\times 2} & 0_{2\times 3} \cr
0_{3\times 2} & V^d_R\cr
\end{array}\right)
\left( \begin{array}{c}  {\hat d}_{Ri} \cr  {\hat b}_{R4} \cr  {\hat b}_{R5} \end{array}\right) + \text{h.c.} \; .
\label{eq:WR}
\end{align}

Note that the matrix $\hat V_{\rm CKM} \equiv U_L^\dagger D_L$ that appears in eq.~(\ref{eq:WL}) is not the standard CKM matrix. In fact, the matrix that controls the SM quark interactions with the $W$ boson is given by:
\begin{align}
V_{\rm CKM} &= \left( \begin{array}{ccccc} 
 \hat V_{ud} &\;&  \hat V_{us} &\;& \hat V_{ub} (V_L^d)_{33} \cr
 \hat V_{cd} &&  \hat V_{cs} && \hat V_{cb} (V_L^d)_{33} \cr
 \hat V_{td} (V_L^u)^\dagger_{33} && \hat V_{ts} (V_L^u)^\dagger_{33}
 && \hat V_{tb} (V_L^u)^\dagger_{33} (V_L^d)_{33} + (V_L^u)^\dagger_{34} (V_L^d)_{43}\cr
  \end{array}\right) \nonumber \\
&  \simeq 
  \left( \begin{array}{ccccc} 
 \hat V_{ud} &\;&  \hat V_{us} &\;& \hat V_{ub} (V_L^d)_{33} \cr
 \hat V_{cd} &&  \hat V_{cs} && \hat V_{cb} (V_L^d)_{33} \cr
 \hat V_{td} (V_L^u)^\dagger_{33} && \hat V_{ts} (V_L^u)^\dagger_{33}
 && \hat V_{tb} (V_L^u)^\dagger_{33} (V_L^d)_{33} \cr
  \end{array}\right)\; ,
\label{eq:CKM}
\end{align} 
where in the last step we used the fact that $(V_L^u)_{33} \simeq 1 - v_u^2 \kappa_T^2/(2 M_T)^2$, $(V_L^d)^\dagger_{33} \simeq 1 - v_d^2 \lambda_B^2/(2 M_B)^2$, while $(V_L^u)^\dagger_{34} (V_L^d)_{43} \sim O(\epsilon^4)$ and thus negligible.

From eqs.~(\ref{eq:WL}) and (\ref{eq:WR}) we can immediately read out the complete set of left and right handed $W$ couplings:
\begin{align}
{\cal L}_W &=  \left( \bar {\hat u}_{La} \gamma^\mu g^{W u_a d_b}_L  \hat d_{Lb} + \bar {\hat u}_{Ra} \gamma^\mu g^{W u_a d_b}_R  \hat d_{Rb} \right) W^+_\mu + h.c. \;, 
\end{align}
where
\begin{align}
g^{W u_i d_j}_L &= \frac{g}{\sqrt{2}} (\hat V_{\rm CKM})_{ij}~~~{\rm for}~~ i,j = 1,2 \; , \\
g^{W u_i d_k}_L &= \frac{g}{\sqrt{2}} (\hat V_{\rm CKM})_{i3} (V_L^d)_{3k}~~~{\rm for}~~ i = 1,2~~\&~~ k = 3, 4, 5 \; , \\
g^{W u_k d_i}_L &= \frac{g}{\sqrt{2}} (\hat V_{\rm CKM})_{3i} (V_L^{u\,\dagger})_{k3}~~~{\rm for}~~ i = 1,2~~\&~~ k = 3, 4, 5 \; , \\
g^{W u_k d_l}_L &= \frac{g}{\sqrt{2}} \left[ (\hat V_{\rm CKM})_{33} (V_L^{u\,\dagger})_{k3} (V_L^d)_{3l} + (V_L^{u\,\dagger})_{k4} (V_L^d)_{4l} \right]~~~{\rm for}~~k,l = 3,4,5  \; , \\
g^{W u_a d_b}_R &= 0~~~{\rm for}~~a = 1,2~~{\rm or}~~b = 1,2 \; , \\
g^{W u_k d_l}_R &= \frac{g}{\sqrt{2}} (V_R^{u\,\dagger})_{k4} (V_R^d)_{4l}~~~{\rm for}~~k,l = 3,4,5 \; . \label{gWtRbR}
\end{align}

\subsection{Couplings of the Z boson}
The couplings of the $Z$ boson to the quarks can be obtained from the kinetic terms:
\begin{align}
{\cal L}_{kin} \supset  \bar u_{La} i \slashed D_a u_{La} +  \bar u_{Ra} i \slashed D_a u_{Ra} + \bar d_{La} i \slashed D_a d_{La} +  \bar d_{Ra} i \slashed D_a d_{Ra}~,
\label{eq:kinN}
\end{align}
where the covariant derivative is 
\begin{eqnarray}
D_{\mu a} = \partial _\mu 
- i \frac{g}{\cos \theta_W} \left( T^3_a  - \sin^2\theta_W Q \right) Z_\mu~, 
\end{eqnarray}
and the weak isospin $T^3_a$ is +1/2 (-1/2) for up (down) components of SU(2) doublets and 0 for singlets. Following the same steps as in the previous section (with the simplification that flavor changing couplings are absent after the rotations in eqs.~(\ref{yudiag}) and (\ref{yddiag})) we obtain the following $Z$ interactions:
\begin{align}
\mathcal{L}_Z &=  \left( \bar f_{La} \gamma^\mu g_L^{Z f_a f_b} f_{Lb} + \bar f_{Ra} \gamma^\mu g_R^{Z f_a f_b} f_{Rb} \right) Z_\mu~, 
\end{align}
where $f_a = \hat u_a$ or $\hat d_a$ and
\begin{align}
g_L^{Z d_i d_j} &= \frac{g}{\cos\theta_W} \left( -\frac12 + \frac13 \sin^2\theta_W \right) \delta_{ij}\,,~~ g_L^{Z d_3 d_j} = 0 ~~~~{\rm for}~~ i,j = 1,2 \; ,\\
g_L^{Z d_k d_l} &= \frac{g}{\cos\theta_W} \left[ \left(- \frac12 + \frac13 \sin^2\theta_W \right) \delta_{kl} + \frac12 (V_L^{d\, \dagger})_{k5} (V_L^d)_{5l}  \right] ~~~~{\rm for}~~ k,l = 3,4,5  \; ,\\
g_R^{Z d_i d_j} &= \frac13 \frac{g}{\cos\theta_W} \sin^2\theta_W \delta_{ij}\,,~~ g_R^{Z d_3 d_j} = 0~~~~{\rm for}~~ i,j = 1,2 \; , \\
g_R^{Z d_k d_l} &= \frac{g}{\cos \theta_W} \left[ \frac13 \sin^2\theta_W \, \delta_{kl} -\frac12 (V_R^{d\, \dagger})_{k4} (V_R^d)_{4l} \right] ~~~~{\rm for}~~ k,l = 3,4,5  \; ,\\
g_L^{Z u_i u_j} &= \frac{g}{\cos\theta_W} \left( \frac12 - \frac23 \sin^2\theta_W \right) \delta_{ij}\,,~~ g_L^{Z u_3 u_j} = 0~~~~{\rm for}~~ i,j = 1,2  \; ,\\
g_L^{Z u_k u_l} &= \frac{g}{\cos\theta_W} \left[ \left(\frac12 - \frac23 \sin^2\theta_W \right) \delta_{kl}  -\frac12 (V_L^{u\, \dagger})_{k5} (V_L^u)_{5l}  \right]~~~~{\rm for}~~ k,l = 3,4,5 \; , \\
g_R^{Z u_i u_j} &= - \frac23 \frac{g}{\cos\theta_W} \sin^2\theta_W \delta_{ij}\,,~~ g_R^{Z u_3 u_j} = 0~~~~{\rm for}~~ i,j = 1,2  \; ,\\
g_R^{Z u_k u_l} &= \frac{g}{\cos \theta_W} \left[ -\frac23 \sin^2\theta_W\, \delta_{kl} + \frac12 (V_R^{u\, \dagger})_{k4} (V_R^u)_{4l} \right] ~~~~ {\rm for}~~ k,l = 3,4,5 \; .
\end{align}
As expected there are no FCNC among the SM quark generations.

\subsection{Couplings of the Higgs bosons}
The couplings of neutral Higgs bosons to up and down quarks follow from the Yukawa interactions in the Lagrangian. The couplings of the first two generations to neutral Higgs bosons are not affected by mixing with the vectorlike fermions. In the basis in which the SM quark Yukawas are diagonal and keeping only the heaviest SM quarks and the vectorlike fermions we have:
\begin{align}
{\cal L}_{H_{u,d}^0}  = \; &
-
\left(\begin{array}{ccc} 
\bar t_{L} & \bar T^Q_{L} & \bar T_{L} 
\end{array}\right) 
\begin{pmatrix}
y_t H_u^0 & 0 & \kappa_T H_u^0  \\
\kappa_Q H_u^0  & 0 & \kappa H_u^0  \\
0 & \bar \kappa H_u^{0\dagger}& 0 \\
\end{pmatrix}
\left(\begin{array}{c} 
t_{R} \cr  T^Q_{R} \cr T_{R} 
\end{array}\right) \nonumber\\  
& -
\left(\begin{array}{ccc} 
\bar b_{L} & \bar B^Q_{L} & \bar B_{L} 
\end{array}\right) 
\begin{pmatrix}
y_b H_d^0  & 0 & \lambda_B H_d^0  \\
\lambda_Q H_d^0  & 0 & \lambda H_d^0  \\
0 & \bar \lambda H_d^{0\dagger} & 0 \\
\end{pmatrix} 
 \left(\begin{array}{c} 
 b_{R} \cr  B^Q_{R} \cr  B_{R} 
\end{array}\right) 
 \; + \; h.c.  \; .
\end{align}
The CP even ($h$ and $H$), CP odd ($A$) and Goldstone boson ($G$) mass eigenstates are:
\begin{align}
\left( \begin{array}{c} H \\ h \\ \end{array} \right)
&=
\left( \begin{array}{cc}
\cos\alpha & \sin\alpha \\
-\sin\alpha & \cos\alpha \\
\end{array} \right)
\left( \begin{array}{c}
\sqrt{2} ( {\rm Re} H_d^0 - v_d )\\
 \sqrt{2} ( {\rm Re} H_u^0 - v_u) \\
\end{array}\right) \; , \\
\left( \begin{array}{c} G \\ A \\ \end{array} \right)
&=
\left( \begin{array}{cc}
\cos\beta & \sin\beta \\
-\sin\beta & \cos\beta \\
\end{array} \right)
\left( \begin{array}{c}
\sqrt{2} ( {\rm Im} H_d^0  )\\
 \sqrt{2} (  - {\rm Im} H_u^0 ) \\
\end{array}\right) \; . 
\label{eq:interactionbasis}
\end{align}
We require that in the CP conserving two Higgs doublet model we consider, the light Higgs $h$ couplings to gauge bosons are identical to those in the SM. This implies that the heavy CP even Higgs $H$ has no couplings to gauge bosons. In this limit, $\alpha = \beta - \pi/2$ and the mass eigenstates $h$ and $H$ read:
\begin{align}
\left( \begin{array}{c} h \\ - H \\ \end{array} \right)
&=
\left( \begin{array}{cc}
\cos\beta & \sin\beta \\
-\sin\beta & \cos\beta \\
\end{array} \right)
\left( \begin{array}{c}
\sqrt{2} ( {\rm Re} H_d^0 - v_d )\\
 \sqrt{2} ( {\rm Re} H_u^0 - v_u) \\
\end{array}\right) \; .
\label{eq:interactionbasis1}
\end{align}
In terms of the mass eigenstates the Lagrangian for $h$ and $H$ reads:
\begin{align}
{\cal L}_{h,H}  = \; &
- \frac{1}{\sqrt{2}}\,
\bar {\hat u}_{L}\, 
V^{u\dagger}_L
Y_u 
V_R^u \, \hat u_{R }
 \, (h\sin\beta  -H\cos\beta )  \nonumber \\
&-\frac{1}{\sqrt{2}}\,
\bar {\hat d}_{L} \,V^{d\dagger}_L
Y_d  
V_R^d \, \hat d_{R}\,
(h\cos\beta +H\sin\beta ) + \; h.c.\, ,
\end{align}
where the $3 \times 3$ matrices $Y_u$ and $Y_d$ are given by 
\begin{eqnarray}
Y_u = 
\begin{pmatrix}
y_t  & 0 & \kappa_T  \\
\kappa_Q  & 0 & \kappa  \\
0 & \bar \kappa & 0 \\
\end{pmatrix} 
\;\;\; \text{and} \;\;\;
Y_d = 
\begin{pmatrix}
y_b  & 0 & \lambda_B  \\
\lambda_Q  & 0 & \lambda  \\
0 & \bar \lambda & 0 \\
\end{pmatrix} \; .
\end{eqnarray}
Since these matrices are not proportional to the corresponding $3\times 3$ minors in the mass matrices given in eqs.~(\ref{eq:mass1}) and (\ref{eq:mass2}), the Higgs couplings are in general flavor violating. The resulting Lagrangian is:
\begin{align}
{\cal L}_{h,H} =\; & 
- \frac{1}{\sqrt{2}} \, \bar {\hat u}_{La}  \, \lambda^h_{u_a u_b} \,  {\hat u}_{Rb}  \, h  
- \frac{1}{\sqrt{2}} \, \bar {\hat d}_{La}  \, \lambda^h_{d_a d_b} \,  {\hat d}_{Rb}  \, h  
\\
\; &  
- \frac{1}{\sqrt{2}} \, \bar {\hat u}_{La}  \, \lambda^H_{u_a u_b} \,  {\hat u}_{Rb}  \, H 
- \frac{1}{\sqrt{2}} \, \bar {\hat d}_{La}  \, \lambda^H_{d_a d_b} \,  {\hat d}_{Rb}  \, H
+h.c. \, ,
\end{align}
where
\begin{align}
\lambda^h_{u_a u_b} &= \sin\beta \, (V_L^{u\,\dagger} Y_u V_R^u)_{ab}~, \label{eq:hu}\\
\lambda^h_{d_a d_b} &= \cos\beta \, (V_L^{d\,\dagger} Y_d V_R^d)_{ab}~, \label{eq:hd}\\
\lambda^H_{u_a u_b} &=  -\cos\beta \, (V_L^{u\,\dagger} Y_u V_R^u)_{ab}~, \label{eq:Hu}\\ 
\lambda^H_{d_a d_b} &=  \sin\beta \, (V_L^{d\,\dagger} Y_d V_R^d)_{ab}~ .\label{eq:Hd}
\end{align}
Since $Y_u v_u = M_t - \text{diag} (0,M_Q,M_T)$, the Higgs boson couplings in the up and down sectors can be written as:
\begin{align}
\lambda^h_{u_a u_b} v &= 
\left(
\begin{array}{ccc}
m_t & 0 & 0 \\
0 & m_{t_4} & 0 \\
0 & 0 & m_{t_5} 
\end{array}
\right) 
- V_L^{u\,\dagger}
\left(
\begin{array}{ccc}
0 & 0 & 0 \\
0 & M_Q & 0 \\
0 & 0 & M_T 
\end{array}
\right) 
V_R^u~, \\
- \lambda^H_{u_a u_b} v \tan\beta &= 
\left(
\begin{array}{ccc}
m_t & 0 & 0 \\
0 & m_{t_4} & 0 \\
0 & 0 & m_{t_5} 
\end{array}
\right) 
- V_L^{u\,\dagger}
\left(
\begin{array}{ccc}
0 & 0 & 0 \\
0 & M_Q & 0 \\
0 & 0 & M_T 
\end{array}
\right) 
V_R^u~, \\
\lambda^h_{d_a d_b} v &= 
\left(
\begin{array}{ccc}
m_b & 0 & 0 \\
0 & m_{b_4} & 0 \\
0 & 0 & m_{b_5} 
\end{array}
\right) 
- V_L^{d\,\dagger}
\left(
\begin{array}{ccc}
0 & 0 & 0 \\
0 & M_Q & 0 \\
0 & 0 & M_B 
\end{array}
\right) 
V_R^d~, \\
\lambda^H_{d_a d_b} \frac{v}{ \tan\beta} &= 
\left(
\begin{array}{ccc}
m_b & 0 & 0 \\
0 & m_{b_4} & 0 \\
0 & 0 & m_{b_5} 
\end{array}
\right) 
- V_L^{d\,\dagger}
\left(
\begin{array}{ccc}
0 & 0 & 0 \\
0 & M_Q & 0 \\
0 & 0 & M_B 
\end{array}
\right) 
V_R^d~, 
\label{eq:huualt}
\end{align}
where we used $v_u = v \sin \beta$ and $v_d = v \cos \beta$. These expressions show explicitly that in the absence of vectorlike fermions the lightest Higgs ($h$) couplings are SM-like, while the heavier scalar Higgs ($H$) couplings in the up (down) sector are suppressed (enhanced) by $\tan \beta$.

The Lagrangian for the CP--odd Higgs $A$ reads:
\begin{align}
{\cal L}_{A}  = \; &
- \frac{i}{\sqrt{2}}\,
\bar {\hat u}_{L}\, 
V^{u\dagger}_L
Y_u^A 
V_R^u \, \hat u_{R }
 \, ( - A \cos\beta )  \nonumber \\
&-\frac{i}{\sqrt{2}}\,
\bar {\hat d}_{L} \,V^{d\dagger}_L
Y_d^A  
V_R^d \, \hat d_{R}\,
(-A \sin\beta ) + \; h.c.\, ,
\end{align}
where the $3 \times 3$ matrices $Y_u^A$ and $Y_d^A$ are given by 
\begin{eqnarray}
Y_u^A = 
\begin{pmatrix}
y_t  & 0 & \kappa_T  \\
\kappa_Q  & 0 & \kappa  \\
0 & -\bar \kappa & 0 \\
\end{pmatrix} 
\;\;\; \text{and} \;\;\;
Y_d^A = 
\begin{pmatrix}
y_b  & 0 & \lambda_B  \\
\lambda_Q  & 0 & \lambda  \\
0 & -\bar \lambda & 0 \\
\end{pmatrix} \; .
\end{eqnarray}
Thus we get:
\begin{align}
{\cal L}_{A} =\; & 
- \frac{1}{\sqrt{2}} \, \bar {\hat u}_{La}  \, \lambda^A_{u_a u_b} \,  {\hat u}_{Rb}  \, A  
- \frac{1}{\sqrt{2}} \, \bar {\hat d}_{La}  \, \lambda^A_{d_a d_b} \,  {\hat d}_{Rb}  \, A  
+h.c. \, ,
\end{align}
where
\begin{align}
\lambda^A_{u_a u_b} &=  - i \cos\beta \, (V_L^{u\,\dagger} Y_u^A V_R^u)_{ab}~, \label{eq:Au}\\
\lambda^A_{d_a d_b} &= -i \sin\beta \, (V_L^{d\,\dagger} Y_d^A V_R^d)_{ab}~. \label{eq:Ad}
\end{align}
Note that in the limit in which there are no CP--violating phases, the left and right components of the diagonal pseudoscalar interactions combine to yield a single coupling proportional to $\gamma^5$. 

Charged Higgs interactions involving SM quarks are controlled by the CKM and can be obtained following the procedure detailed in section~\ref{sec:wboson}. We focus only on interactions involving the third generation of SM quarks and the heavy vectorlike fermions for which CKM effects are negligible. In the basis in which the SM quark Yukawas are diagonal and keeping only the heaviest SM quarks and the vectorlike fermions we have:
\begin{align}
{\cal L}_{H_{u,d}^\pm}  = \; & 
- 
\left(\begin{array}{ccc} 
\bar t_{L} & \bar T^Q_{L} & \bar T_{L} 
\end{array}\right) 
\begin{pmatrix}
y_b H_d^+ & 0 & \lambda_B H_d^+  \\
\lambda_Q H_d^+  & 0 & \lambda H_d^+  \\
0 & \bar \kappa H_u^+& 0 \\
\end{pmatrix}
\left(\begin{array}{c} 
b_{R} \cr  B^Q_{R} \cr B_{R} 
\end{array}\right) \nonumber\\  
& -
\left(\begin{array}{ccc} 
\bar b_{L} & \bar B^Q_{L} & \bar B_{L} 
\end{array}\right) 
\begin{pmatrix}
y_t H_u^-  & 0 & \kappa_T H_u^-  \\
\kappa_Q H_u^-  & 0 & \kappa H_u^-  \\
0 & \bar \lambda H_d^- & 0 \\
\end{pmatrix} 
 \left(\begin{array}{c} 
 t_{R} \cr  T^Q_{R} \cr  T_{R} 
\end{array}\right) 
 \; + \; h.c.  \; ,
\end{align}
where we adopt the conventional notation $H_d^- \equiv (H_d^+)^\dagger$ and $H_u^+ \equiv (H_u^-)^\dagger$. The charged Higgs ($H^\pm$) and Goldstone bosons ($G^\pm$) mass eigenstates are given by:
\begin{align}
\left( \begin{array}{c} G^\pm \\H^\pm \\ \end{array}\right)
&=
\left( \begin{array}{cc}
\cos\beta & \sin\beta \\
-\sin\beta & \cos\beta \\
\end{array}\right)
\left( \begin{array}{c}
H_d^\pm  \\
 - H_u^\pm \\
\end{array}\right) \; .
\label{eq:chargedHiggsbasis}
\end{align}
In terms of the mass eigenstates the Lagrangian for $H^\pm$ reads:
\begin{align}
{\cal L}_{H^\pm}  = \; &
- 
\bar {\hat u}_{L}\, 
V^{u\dagger}_L
Y_d^{H^\pm} 
V_R^d \, \hat d_{R }
 \, H^+  \nonumber \\
&-
\bar {\hat d}_{L} \,V^{d\dagger}_L
Y_u^{H^\pm}  
V_R^u \, \hat u_{R}\,
H^- + \; h.c.\, ,
\end{align}
where the $3 \times 3$ matrices $Y_u^{H^\pm} $ and $Y_d^{H^\pm} $ are given by 
\begin{eqnarray}
Y_u^{H^\pm} =
 -  
\cos\beta 
\begin{pmatrix}
y_t  & 0 & \kappa_T  \\
\kappa_Q  & 0 & \kappa  \\
0 & \bar \lambda\tan\beta  & 0 \\
\end{pmatrix} 
\;\;\; \text{and} \;\;\;
Y_d^{H^\pm} = 
-\sin\beta 
\begin{pmatrix}
y_b  & 0 & \lambda_B  \\
\lambda_Q  & 0 & \lambda  \\
0 & \bar \kappa/\tan\beta & 0 \\
\end{pmatrix} \; .
\end{eqnarray}
Thus we get:
\begin{align}
{\cal L}_{H^\pm} =\; & 
- \bar {\hat u}_{La}  \, \lambda^{H^\pm}_{u_a d_b} \,  {\hat d}_{Rb}  \, H^+  
-  \bar {\hat d}_{La}  \, \lambda^{H^\pm}_{d_a u_b} \,  {\hat u}_{Rb}  \, H^- 
+h.c. \; ,
\end{align}
where 
\begin{align}
\lambda^{H^\pm}_{u_a d_b} &= (V_L^{u\,\dagger} Y_d^{H^\pm} V_R^d)_{ab}~, \\
\lambda^{H^\pm}_{d_a u_b} &= (V_L^{d\,\dagger} Y_u^{H^\pm} V_R^u)_{ab}~. 
\label{eq:hcc}
\end{align}
Explicit expressions for the couplings relevant to our analysis are:
\begin{align}
\lambda^{H^\pm}_{t_4 b} &= -\sin\beta \cdot \left[ y_b (V_L^u)^\dagger_{43} (V_R^d)_{33} + \lambda_B (V_L^u)^\dagger_{43} (V_R^d)_{53} + \lambda_Q (V_L^u)^\dagger_{44} (V_R^d)_{33}  + \lambda (V_L^u)^\dagger_{44} (V_R^d)_{53} \right] \nonumber \\
&\hspace{1.5cm}  - \cos\beta   \bar \kappa (V_L^u)^\dagger_{45} (V_R^d)_{43}
\,,\label{eq:charged1}\\
\lambda^{H^\pm}_{b t_4} &=  - \cos\beta \cdot \left[ 
y_t (V_L^d)^\dagger_{33} (V_R^u)_{34} +\kappa_T (V_L^d)^\dagger_{33} (V_R^u)_{54} + \kappa_Q (V_L^d)^\dagger_{34} (V_R^u)_{34} + \kappa (V_L^d)^\dagger_{34} (V_R^u)_{54} \right] \nonumber \\
&\hspace{1.5cm} -\sin\beta \cdot \bar \lambda (V_L^d)^\dagger_{35} (V_R^u)_{44} \,, \\
\lambda^{H^\pm}_{b_4 t} &=  - \cos\beta \cdot \left[ 
y_t (V_L^d)^\dagger_{43} (V_R^u)_{33} +
\kappa_T (V_L^d)^\dagger_{43} (V_R^u)_{53} + \kappa_Q (V_L^d)^\dagger_{44} (V_R^u)_{33} + \kappa (V_L^d)^\dagger_{44} (V_R^u)_{53} \right] \nonumber \\
&\hspace{1.5cm} -\sin\beta \cdot \bar \lambda (V_L^d)^\dagger_{45} (V_R^u)_{43} \,, \\
\lambda^{H^\pm}_{t b_4} &= -\sin\beta \cdot \left[ y_b (V_L^u)^\dagger_{33} (V_R^d)_{34} + \lambda_B (V_L^u)^\dagger_{33} (V_R^d)_{54} + \lambda_Q (V_L^u)^\dagger_{34} (V_R^d)_{34}  + \lambda (V_L^u)^\dagger_{34} (V_R^d)_{54} \right] \nonumber \\
&\hspace{1.5cm}  - \cos\beta  \bar \kappa (V_L^u)^\dagger_{35} (V_R^d)_{44}
 \,.\label{eq:charged4}
\end{align}

\subsection{Partial decay widths of vectorlike quarks into the $W$, $Z$, $h$, $H$ and $H^\pm$}
\label{sec:VLQdecay}
In this section, we present explicit expressions for the partial widths of the vectorlike quarks. The partial width for $t_i \to W b_j$ for $i,j = 3,4,5$, if kinematically allowed, is given by:
\begin{align}
\Gamma(t_i \to W b_j) &= \frac{m_{t_i}}{32 \pi} 
\sqrt{\lambda\left(1, \frac{M_W^2}{m_{t_i}^2},\frac{m_{b_j}^2}{m_{t_i}^2}\right)}
\nonumber \\
&\hspace{0.5cm} \times \left\{ \left( \left| g_L^{W t_i b_j} \right|^2 + \left| g_R^{W t_i b_j} \right|^2 \right) \left( 1 + \frac{m_{t_i}^2 - 2 m_{b_j}^2}{M_W^2} + \frac{m_{b_j}^2 - 2 M_W^2}{m_{t_i}^2} + \frac{m_{b_j}^4}{M_W^2 m_{t_i}^2} \right) \right.  \nonumber \\
&\hspace{1.5cm} \left.  - \,6 \left(  (g_L^{W t_i b_j})^\ast g_R^{W t_i b_j} + g_L^{W t_i b_j} (g_R^{W t_i b_j})^\ast \right) \frac{m_{b_j}}{m_{t_i}} \right\}\;.
\label{eq:decayt4W}
\end{align}
The decay width for $b_i \to W t_j$ can be obtained by replacing $t_i \to b_i$ and $b_j \to t_j$. Similarly, the partial width for $t_i \to Z t_j$ for $i,j = 3,4,5$ is given by:
\begin{align}
\Gamma(t_i \to Z t_j) &= \frac{m_{t_i}}{32 \pi} 
\sqrt{\lambda\left(1, \frac{M_Z^2}{m_{t_i}^2},\frac{m_{t_j}^2}{m_{t_i}^2}\right)} 
 \nonumber \\
&\hspace{0.5cm} \times \left\{ \left( \left| g_L^{Z t_i t_j} \right|^2 + \left| g_R^{Z t_i t_j} \right|^2 \right)  \left( 1 + \frac{m_{t_i}^2 - 2 m_{t_j}^2}{M_Z^2} + \frac{m_{t_j}^2 - 2 M_Z^2}{m_{t_i}^2} + \frac{m_{t_j}^4}{M_Z^2 m_{t_i}^2} \right) \right.   \nonumber \\
&\hspace{2cm} \left.  - \,6  \left( (g_L^{Z t_i t_j})^\ast g_R^{Z t_i t_j} + g_L^{Z t_i t_j} (g_R^{Z t_i t_j})^\ast \right) \frac{m_{t_j}}{m_{t_i}} \right\}  \;.
\label{eq:decayt4Z}
\end{align}
The decay width for $b_i \to Z b_j$ can be obtained by replacing $t_i \to b_i$ and $t_j \to b_j$. The partial width for $t_i \to h t_j$ for $i,j=3,4,5$ is given by:
\begin{align}
\Gamma(t_i \to h t_j) &= \frac{m_{t_i}}{64 \pi} 
\sqrt{\lambda\left(1, \frac{m_h^2}{m_{t_i}^2},\frac{m_{t_j}^2}{m_{t_i}^2}\right)}
\left\{ \left( \left| \lambda^h_{t_i t_j} \right|^2  + \left| \lambda^h_{t_j t_i} \right|^2 \right) \left( 1 + \frac{m_{t_j}^2 - m_h^2}{m_{t_i}^2} \right) \right.
\nonumber \\
&\hspace{0.2cm} \left. + 2 \left( (\lambda^h_{t_i t_j})^\ast \lambda^h_{t_j t_i} + \lambda^h_{t_i t_j} (\lambda^h_{t_j t_i})^\ast \right)\frac{m_{t_j}}{m_{t_i}} \right\}  \; .
\label{eq:decayt4h}
\end{align}
The decay width for  $b_i \to h b_j$ can be obtained by replacing $t_i \to b_i$ and $t_j \to b_j$. The formulas presented above are consistent with those in  ref.~\cite{Atre:2011ae}.

Finally, the decay widths into heavy neutral and charged Higgs bosons are given by:
\begin{align}
\Gamma(t_i \to H t_j) &= \frac{m_{t_i}}{64 \pi}  
\sqrt{\lambda\left(1, \frac{m_H^2}{m_{t_i}^2},\frac{m_{t_j}^2}{m_{t_i}^2}\right)}
\left\{ \left( \left| \lambda^H_{t_i t_j} \right|^2  + \left| \lambda^H_{t_j t_i} \right|^2 \right) \left( 1 + \frac{m_{t_j}^2 - m_H^2}{m_{t_i}^2} \right) \right.
\nonumber \\
&\hspace{0.2cm} \left. + 2 \left( (\lambda^H_{t_i t_j})^\ast \lambda^H_{t_j t_i} + \lambda^H_{t_i t_j} (\lambda^H_{t_j t_i})^\ast \right)\frac{m_{t_j}}{m_{t_i}} \right\}  \; , \\
\Gamma(t_i \to H^\pm b_j) &= \frac{m_{t_i}}{ 32 \pi}  
\sqrt{\lambda\left(1, \frac{m_{H^\pm}^2}{m_{t_i}^2},\frac{m_{b_j}^2}{m_{t_i}^2}\right)}
\left\{ \left( \left| \lambda^{H^\pm}_{t_i b_j} \right|^2  + \left| \lambda^{H^\pm}_{b_j t_i} \right|^2 \right) \left( 1 + \frac{m_{b_j}^2 - m_{H^\pm}^2}{m_{t_i}^2} \right) \right.
\nonumber \\
&\hspace{0.2cm} \left.  + 2 \left( (\lambda^{H^\pm}_{t_i b_j})^\ast \lambda^{H^\pm}_{b_j t_i} + \lambda^{H^\pm}_{t_i b_j} (\lambda^{H^\pm}_{b_j t_i})^\ast \right)\frac{m_{b_j}}{m_{t_i}} \right\}  \; .
\end{align}

\subsection{Off-diagonal couplings in special cases}
\label{app:special}
In this subsection, we present approximated formulas for the off-diagonal Yukawa and gauge couplings between a vectorlike heavy quark $t_j$ (where $j = 4, 5$) and the third generation SM quarks in special cases where $t_j$ is almost a singlet or a doublet. We assume that all $\kappa$'s and $\lambda$'s are small parameters of the same order of magnitude. The mixing matrices given in eqs.~(\ref{eq:VLu})--(\ref{eq:VRd}) depend on the four combinations:
\begin{align}
\epsilon_{Q,T}^u &\equiv \frac{v_u}{M_{Q,T}} \times (\kappa_Q, \kappa_T, \kappa, \bar \kappa)\,, \\
\epsilon_{Q,B}^d &\equiv \frac{v_d}{M_{Q,B}} \times (\lambda_Q, \lambda_B, \lambda, \bar \lambda)\,,
\end{align}
where $\epsilon_Q^{u,d} \gg \epsilon_{T,B}^{u,d}$ ($\epsilon_Q^{u,d} \ll \epsilon_{T,B}^{u,d}$) for a $SU(2)$ doublet (singlet) vectorlike quark (this is simply due to the fact that $m_{t_4,b_4} \sim M_Q \ll M_T$ for a $SU(2)$ doublet and $m_{t_4,b_4} \sim M_T \ll M_Q$ for a $SU(2)$ singlet. We remind the reader that eqs.~(\ref{eq:VLu})--(\ref{eq:VRd}) have been derived under the assumption that the lightest vectorlike quark is a doublet; the singlet case is obtained by swapping the second and third column in each matrix.

For almost doublet $t_j$ ($\epsilon_Q^u \gg \epsilon_T^u$) the off-diagonal Yukawa and gauge couplings read:
\begin{align}
\lambda^h_{t t_j} &= - \frac{M_Q}{v} (V^u_L)^\dagger_{34} (V^u_R)_{4j} - \frac{M_T}{v}  (V_L^u)^\dagger_{35} (V^u_R)_{5j}  \nonumber \\
&\simeq  v_u \sin\beta \left( \frac{y_t \kappa_Q}{M_Q} - \frac{\kappa_T \bar \kappa}{M_T} - \frac{\kappa_T \kappa M_Q + \kappa_T \bar \kappa M_T}{M_T^2 - M_Q^2} \right) \nonumber \\
&
\hskip -0.5cm \stackrel{M_Q \ll M_T}{\simeq}  \left[  - 2 \frac{v_u}{M_T}\kappa_T \bar \kappa \sin\beta+\frac{v_u}{M_Q} y_t  \kappa_Q \sin\beta - \frac{v_u M_Q}{M_T^2} \kappa_T \kappa \sin\beta \right] \,, 
\label{eq:doubletlam34} \\
\lambda^h_{t_j t} &= - \frac{M_Q}{v} (V^u_L)^\dagger_{j4} (V^u_R)_{43} - \frac{M_T}{v}  (V_L^u)^\dagger_{j5} (V^u_R)_{53} \nonumber \\
&\simeq \kappa_Q \sin\beta + v_u^2 \sin\beta \left( \frac{\bar \kappa M_Q + \kappa M_T}{M_T^2 - M_Q^2} \right) \left( \frac{\kappa_Q \bar \kappa}{M_Q} - \frac{y_t \kappa_T}{M_T} \right) \nonumber \\
&
\hskip -0.5cm \stackrel{M_Q \ll M_T}{\simeq} 
\left[ \kappa_Q \sin\beta + v_u^2 \sin\beta \left( \frac{\kappa \kappa_Q \bar \kappa}{M_Q M_T}  + \frac{\bar \kappa^2 \kappa_Q -y_t \kappa \kappa_T}{M_T^2} - \frac{y_t \bar \kappa \kappa_T M_Q}{M_T^3} \right) \right]\,,
\label{eq:doubletlam43} \\
g_L^{W t_j b} &= \frac{g}{\sqrt{2}} \left[ (V_{\rm CKM})_{33} (V^u_L)^\dagger_{j3} (V^d_L)_{33} + (V^u_L)^\dagger_{j4} (V^d_L)_{43} \right]\, \nonumber \\
&\simeq - \frac{g}{\sqrt{2}}  v_u^2 \left\{ \frac{\kappa_T}{M_Q} \left( \frac{\kappa M_Q + \bar \kappa M_T}{M_T^2 - M_Q^2} \right)  - \frac{y_t \kappa_Q}{M_Q^2} \right\}  + \frac{g}{\sqrt{2}} v_d^2 \left( \frac{\lambda_B \bar \lambda}{M_Q M_B} - \frac{y_b \lambda_Q}{M_Q^2} \right) \nonumber \\
&
\hskip -0.5cm \stackrel{M_Q \ll M_T}{\simeq}
-\frac{g}{\sqrt{2}} \frac{v}{M_Q} \Bigg[ 
 \frac{v_u}{M_T} \kappa_T \bar \kappa\sin\beta
- \frac{v_u}{M_Q}  y_t \kappa_Q \sin\beta + \frac{v_d}{M_Q}  y_b \lambda_Q \cot\beta  \nonumber \\
& -\frac{v_d}{M_B}\lambda_B \bar \lambda \cot\beta  +  \frac{v_u M_Q }{M_T^2} \kappa \kappa_T\sin\beta\Bigg] \,, 
\label{eq:doubletgLW43} \\
g_R^{W t_j b} &= \frac{g}{\sqrt 2} (V^u_R)^\dagger_{j4} (V^d_R)_{43} \nonumber \\
&
\hskip -0.5cm \stackrel{M_Q \ll M_T}{\simeq}
- \frac{g}{\sqrt 2} \frac{v}{M_Q} \left[ - \lambda_Q \cos\beta \right]  \,, 
\label{eq:doubletgRW43} \\
g_L^{Z t_j t} &= - \frac{g}{2 \cos\theta_W} (V^u_L)^\dagger_{j5} (V^u_L)_{53} \nonumber \\
&\simeq - \frac{g}{2 \cos\theta_W} \left( - v_u \frac{\bar \kappa M_Q + \kappa M_T}{M_T^2 - M_Q^2} \right) \left( -v_u \frac{\kappa_T}{M_T} \right) \nonumber \\
&
\hskip -0.5cm \stackrel{M_Q \ll M_T}{\simeq}
  - \frac{g}{2 \cos\theta_W} \frac{v}{M_Q} \left[ \frac{v_u M_Q }{M_T^2} \kappa \kappa_T\sin\beta + \frac{ v_u M_Q^2}{M_T^3} \bar \kappa \kappa_T \sin\beta\right]  \,,
\label{eq:doubletgLZ43} \\
g_R^{Z t_j t} &= \frac{g}{2 \cos\theta_W} (V^u_R)^\dagger_{j4} (V^u_R)_{43} \nonumber \\
&
\hskip -0.5cm \stackrel{M_Q \ll M_T}{\simeq}
- \frac{g}{2 \cos\theta_W} \frac{v }{M_Q} \left[ \kappa_Q \sin\beta\right] \,,
\label{eq:doubletgRZ43}
\end{align}
up to $\mathcal O ((\epsilon_Q^u)^3)$.

For almost singlet $t_j$ ($\epsilon_T^u \gg \epsilon_Q^u$) the off-diagonal Yukawa and gauge couplings read:
\begin{align}
\lambda^h_{t t_j} &= - \frac{M_Q}{v} (V^u_L)^\dagger_{34} (V^u_R)_{4j} - \frac{M_T}{v}  (V_L^u)^\dagger_{35} (V^u_R)_{5j}  \nonumber \\
&\simeq \kappa_T \sin\beta - v_u^2 \sin\beta \left(  \frac{\kappa_T \bar \kappa}{M_T} - \frac{y_t \kappa_Q}{M_Q}  \right) \left(   \frac{\kappa M_Q + \bar \kappa M_T}{M_T^2 - M_Q^2} \right) \nonumber \\
& \hskip -0.5cm \stackrel{M_T \ll M_Q}{\simeq}
\left[
\kappa_T \sin\beta + v_u^2 \sin\beta \left( \frac{\kappa \kappa_T \bar \kappa}{M_Q M_T} + \frac{\kappa_T \bar \kappa^2 - y_t \kappa \kappa_Q}{M_Q^2} - \frac{y_t \kappa_Q \bar \kappa M_T}{M_Q^3}   \right) \right] \,,  
\label{eq:singletlam34} \\
\lambda^h_{t_j t} &= - \frac{M_Q}{v} (V^u_L)^\dagger_{j4} (V^u_R)_{43} - \frac{M_T}{v}  (V_L^u)^\dagger_{j5} (V^u_R)_{53} \nonumber \\
&\simeq v_u \sin\beta  \left( \frac{\kappa_Q \bar \kappa M_Q + \kappa_Q \kappa M_T}{M_T^2 - M_Q^2}  + \frac{y_t \kappa_T}{M_T} - \frac{\kappa_Q \bar \kappa}{M_Q}  \right)   \nonumber \\
& \hskip -0.5cm \stackrel{M_T \ll M_Q}{\simeq}
 \left[ v_u \sin\beta \left( \frac{y_t \kappa_T}{M_T} - \frac{2 \bar \kappa \kappa_Q}{M_Q} - \frac{\kappa \kappa_Q M_T}{M_Q^2}   \right) \right]\,, 
\label{eq:singletlam43} \\
g_L^{W t_j b} &= \frac{g}{\sqrt{2}} \left[ (V_{\rm CKM})_{33} (V^u_L)^\dagger_{j3} (V^d_L)_{33} + (V^u_L)^\dagger_{j4} (V^d_L)_{43} \right] \nonumber \\
&\simeq  \frac{g}{\sqrt{2}}  \left[ v_u \frac{\kappa_T}{M_T} + v_u \left( \frac{\bar \kappa M_Q + \kappa M_T}{M_T^2 - M_Q^2} \right)  v_d^2 \left( \frac{\lambda_B \bar \lambda}{M_Q M_B} - \frac{y_b \lambda_Q}{M_Q^2} \right) \right]\nonumber \\
&\simeq  \frac{g}{\sqrt{2}} \frac{v}{M_T} \left[ \kappa_T \sin\beta \right] \,,
\label{eq:singletgLW43} \\
g_R^{W t_j b} &= \frac{g}{\sqrt 2} (V^u_R)^\dagger_{j4} (V^d_R)_{43} \nonumber \\
&\simeq \frac{g}{\sqrt 2} v_u \left(  \frac{\kappa M_Q + \bar \kappa M_T}{M_T^2 - M_Q^2} \right) \left( - v_d \frac{\lambda_Q}{M_Q} \right) \nonumber \\
& \hskip -0.5cm \stackrel{M_T \ll M_Q}{\simeq}
 \frac{g}{\sqrt 2} \frac{v}{M_T} \left[ \left( \kappa\frac{ v_d  M_T}{M_Q^2} + \bar \kappa \frac{ v_d M_T^2}{M_Q^3} \right) \lambda_Q \sin\beta\right]\,, 
\label{eq:singletgRW43} \\
g_L^{Z t_j t} &= - \frac{g}{2 \cos\theta_W} (V^u_L)^\dagger_{j5} (V^u_L)_{53} \simeq  \frac{g}{2 \cos\theta_W} \frac{v}{M_T} \left[\kappa_T \sin\beta\right]\,, 
\label{eq:singletgLZ43} \\
g_R^{Z t_j t} &= \frac{g}{2 \cos\theta_W} (V^u_R)^\dagger_{j4} (V^u_R)_{43} \nonumber \\
&\simeq \frac{g}{2 \cos\theta_W} v_u \left(  \frac{\kappa M_Q + \bar \kappa M_T}{M_T^2 - M_Q^2} \right) \left( - \frac{v_u \kappa_Q}{M_Q} \right) \nonumber \\
& \hskip -0.5cm \stackrel{M_T \ll M_Q}{\simeq}
\frac{g}{2 \cos\theta_W} \frac{v}{M_T} \left[\kappa \kappa_Q \sin\beta \frac{v_u M_T}{M_Q^2} + \bar \kappa \kappa_Q \sin\beta \frac{v_u M_T^2}{M_Q^3} \right]\,,
\label{eq:singletgRZ43} 
\end{align}
up to $\mathcal O ((\epsilon_T^u)^3)$.

Couplings of $b_j$ are obtained by replacing $\kappa\text{'s} \leftrightarrow \lambda\text{'s}$, $y_t\leftrightarrow y_b$, and $\beta \to \beta + \pi/2$.

Note that in the case of a singlet heavy vectorlike quark in absence of any mixing with a doublet, there are only left-handed couplings to $W$, $Z$ and $h$ and we have $g^W \simeq \frac{g}{\sqrt{2}}\frac{v}{M} \lambda^h$ and $g^Z \simeq \frac{g}{2\cos\theta_W}\frac{v}{M} \lambda^h$. Then eqs.~(\ref{eq:decayt4W})-(\ref{eq:decayt4h}) directly lead to $\Gamma_W/2 = \Gamma_Z = \Gamma_h  = M/(64 \pi) |\lambda_h|^2$, implying that the $W$, $Z$ and $h$ branching ratios are 50\%, 25\% and 25\%. 

In the case of a heavy $SU(2)$ doublet vectorlike quark in absence of any mixing with a singlet, there are two relevant parameters which control the couplings to $W$, $Z$ and $h$. While the latter two are related as in the singlet case, $g^W$ is independent; implying that the branching ratio into $W$ is arbitrary while the $Z$ and $h$ channels are identical.

\subsection{On ellipses}
\label{ellipses}
Here we show that if a vectorlike quark decays into three channels ($Z$, $W$, $h$) whose effective couplings are linear combinations of two fundamental parameters $a$ and $b$, any two branching ratios lie on an ellipse. By assumption the widths $\Gamma_{W,Z,h}$ are linear combinations of $a^2$, $b^2$ and $ab$.
A generic ellipse in the $[{\rm BR}_Z,{\rm BR}_W]$ plane is given by the equation:
\begin{align}
0 &= \epsilon_1 {\rm BR}_Z^2 + \epsilon_2 {\rm BR}_W^2 + \epsilon_3 {\rm BR}_Z {\rm BR}_W + \epsilon_4 {\rm BR}_Z + \epsilon_5 {\rm BR}_W + 1 \\
&= \frac{\text{linear combination of $a^4$, $b^4$, $a^2b^2$, $a^3b$ and $ab^3$}}{(\Gamma_Z+\Gamma_W+\Gamma_h)^2} \;,
\end{align}
which admits a unique solution for the five coefficients $\epsilon_i$.

\end{document}